\documentclass{emulateapj}
\slugcomment{{\sc Accepted to Astronomy and Computing:} April 2016}
\usepackage{amsmath} 
\usepackage{float}
\usepackage{natbib} \usepackage{hyperref}

\newcommand{\oxabinline}{\ensuremath{12 + \log_{10}(O/H)}}
\newcommand{\ha}{\ensuremath{\mathrm{H}\alpha}}
\newcommand{\hb}{\ensuremath{\mathrm{H}\beta}}
\newcommand{\ebmv}{\ensuremath{\mathrm{\emph{E(B-V)}}}}

\begin{document}
\title{Monte Carlo Method for Calculating Oxygen Abundances and Their
  Uncertainties from Strong-Line Flux Measurements}

\author{Federica B. Bianco\altaffilmark{1,2}, Maryam
  Modjaz\altaffilmark{1}, Seung Man Oh\altaffilmark{1,3}, David
  Fierroz\altaffilmark{1}, Yuqian Liu\altaffilmark{1}, Lisa
  Kewley\altaffilmark{4,5}, Or Graur\altaffilmark{1,6}}
\altaffiltext{1}{Center for Cosmology and Particle Physics, New York
  University, 4 Washington Place, New York, NY 10003, USA}
\altaffiltext{2}{Center for Urban Science and Progress, New York
  University, 1 MetroTech Center, Brooklyn, NY 11201}
\altaffiltext{3}{NYU Abu Dhabi PO Box 129188 Abu Dhabi, UAE}
\altaffiltext{4}{Australian National University, Research School for
  Astronomy \& Astrophysics, Mount Stromlo Observatory, Cotter Road,
  Weston, ACT 2611, Australia } \altaffiltext{5}{ Institute of
  Astronomy, University of Hawaii, 2680 Woodlawn Drive, Honolulu, HI
  96822, USA} \altaffiltext{6}{Department of Astrophysics, American
  Museum of Natural History, Central Park West and 79th Street, New
  York, NY 10024-5192, USA.}

\begin{abstract}
We present the open-source Python code \verb=pyMCZ= that determines
oxygen abundance and its distribution from strong emission lines in
the standard metallicity calibrators, based on the original IDL code
of \citet{kewley02} with updates from \citet{kewley08}, and expanded
to include more recently developed calibrators.  The standard
strong-line diagnostics have been used to estimate the oxygen
abundance in the interstellar medium through various emission line
ratios (referred to as indicators) in many areas of astrophysics,
including galaxy evolution and supernova host galaxy studies.  We
introduce a Python implementation of these methods that, through Monte
Carlo sampling, better characterizes the statistical oxygen abundance
confidence region including the effect due to the propagation of
observational uncertainties.  These uncertainties are likely to
dominate the error budget in the case of distant galaxies, hosts of
cosmic explosions.  Given line flux measurements and their
uncertainties, our code produces synthetic distributions for the
oxygen abundance in up to 15 metallicity calibrators simultaneously,
as well as for \ebmv, and estimates their median values and their 68\%
confidence regions.  We provide the option of outputting the full
Monte Carlo distributions, and their Kernel Density estimates.  We
test our code on emission line measurements from a sample of nearby
supernova host galaxies ($z<0.15$) and compare our metallicity results
with those from previous methods. We show that our metallicity
estimates are consistent with previous methods but yield smaller
statistical uncertainties. It should be noted that systematic
uncertainties are not taken into account.  We also offer visualization
tools to assess the spread of the oxygen abundance in the different
calibrators, as well as the shape of the estimated oxygen abundance
distribution in each calibrator, and develop robust metrics for
determining the appropriate Monte Carlo sample size.  The code is open
access and open source and can be found at
\url{https://github.com/nyusngroup/pyMCZ}
\end{abstract}

\section{Introduction}
Small amounts of carbon, oxygen, nitrogen, sulfur, iron and other
elements provide a splash of color to the otherwise dominating
grayscape of hydrogen and helium in the stars and gas of
galaxies. Nevertheless, even this minute presence of heavy elements is
important for many areas of astrophysics. For example,
\citet{johnson12} suggest that if it were not for the relatively high
metallicity level in our Solar System, planet formation may not have
been possible. With $Z$ representing the mass fraction of metals, all
elements heavier than He (also collectively called metallicity), for
our own Sun the value is measured to be $Z=0.0153$ \citep{chaffau11},
though there are suggestions of a lower Solar metallicity of
$Z=0.0134$ \citep{asplund09_rev,grevesse10}.  When properly observed
and estimated, metallicity measurements of galaxies can tightly
constrain models of galaxy formation and evolution (e.g.,
\citealt{kewley08} and references therein), and shed light on the
metallicity dependence and production conditions for different types
of supernovae (SNe) and long-duration gamma-ray bursts (GRBs) (e.g.,
\citealt{modjaz08_Z,levesque10_grbhosts,anderson10, modjaz11, kelly12,
  sanders12, leloudas14, lunnan14, pan14}).

Metals are produced in the cores of massive stars during their fusion
life cycle but also during the extreme conditions of stellar
explosions. For example, the majority of iron is synthesized in
thermonuclear explosions (SNe Ia) while nearly all oxygen and other
$\alpha$-elements\footnote{$\alpha$-elements are element with even
  atomic number lower than 22, synthesized by $\alpha$-capture} are
released in various kinds of core-collapse SNe (SNe II and
stripped-envelope core-collapse SNe).

However, for almost all astronomical objects, metallicity cannot be
measured directly. The oxygen abundance in the gas-phase is the
canonical choice of metallicity indicator for interstellar medium
(ISM) studies. After Hydrogen, Oxygen lines are the strongest emission
lines in optical wavelengths. Oxygen is the most abundant metal and it
is only weakly depleted onto dust grains, in contrast to refractory
elements such as magnesium, silicon, or iron (iron for example is
depleted by more than a factor of 10 in Orion; see
\citealt{simondiaz11-orion}). The oxygen abundance is expressed as
\oxabinline, where O and H represent the number of oxygen and hydrogen
atoms, respectively. For example: \citealt{chaffau11} measure a Solar
oxygen abundance of \oxabinline = 8.76 $\pm$ 0.07, while
\citet{asplund09_rev} suggest \oxabinline = 8.69.\footnote{The ongoing
  debate about the value of the Solar oxygen abundance should be kept
  in mind when metallicities are expressed relative to solar
  metallicity.}  While oxygen abundance is used to gauge metallicity,
in many cases in the literature, including here, the terms metallicity
and oxygen abundance are used interchangeably.  Importantly, oxygen
exhibits very strong nebular lines in the optical wavelength range in
spectra of HII regions (e.g.,
\citealt{pagel79,osterbrock89,tremonti04}), which can be
measured. Thus, many different diagnostic techniques relying on
different lines of oxygen, hydrogen and other elements, have been
developed (e.g., \citealt{kewley02} -- hereafter KD02,
\citealt{pettini04} -- PP04, \citealt{kobulnicky04} -- KK04,
\citealt{kewley08} -- KE08), and are discussed in the next section.

Many fields rely on the determination of the metallicity environments
to understand physical phenomena, such as SN and planetary formation,
from a causal, and possibly mechanicistic point of view. However many
of these fields, including SN studies, have used metallicity
calibration techniques somewhat acritically, at times inferring from
calibrators that are not comparable, as they rely on different
assumptions, because of the scarsity of complete dataset that would
allow consistent estimates, and often ignoring many sources of
uncertainty.

The purpose of this paper is to present a public code that collects
different abundance diagnostics to efficiently and rapidly compute
metallicity from strong emission line fluxes as well as the associated
statistical uncertainties due to the measured emission line flux
uncertainties. This source of uncertainty is particularly relevant in
the study of SN metallicity environments, since the host galaxies of
these cosmics explosions are often at large distances. While in many
other contexts the signal-to-noise (SNR) of the spectra themselves may
contribute negligebly to the total uncertainty budget, compared for
example to errors in the model parameters, it is often a substantial
source of uncertainty for high $Z$ galaxies that host SNe.  The
necessity to compute metallicity quickly, and systematically for large
samples arises from the new availability of large samples, enabled by
IMACS spectrographs for example, and collected by surveys such as
MANGA \citep{Bundy15}.

Our open-source Python package is named \textbf{pyMCZ}.  This Python
code allows the user to quickly produce metallicity values with
sensible confidence regions for several metallicity calibrators at
once, given an input set of spectral line measurements and their
errors, and to obtain and visualize the distribution of metallicities.
\verb=pyMCZ= is structured in a modular way in order to allows the
user to include other strong line metallicity calibration methods, and
link to external code packages, with minimal modifications, and
naturally extend the advantages of the Monte Carlo (MC) based
propagation of the observational uncertainties to the newly included
calibrators.  While we do not mean to advocate for a particular
metallicity calibrator to be adopted, the comparison of multiple
calibrator outputs, and the shape of each metallicity distribution,
can guide the user in understanding the reliability of a metallicity
estimate, given a set of line fluxes.  Below we describe the different
oxygen abundance diagnostics, and our Python module implementation.

\section{Oxygen abundance diagnostics and calibrators}\label{sec:diags}

In this section we present a brief overview of the various
observational methods for measuring the gas-phase oxygen abundance -
however, for a detailed discussion, and to understand the many
caveats, we encourage the reader to read the reviews by, e.g.,
\citet{stasinska02}, KE08, \citet{moustakas10}, \citet{stasinska10},
\citet{lopezsanchez12}, \citet{dopita13}, hereafter D13, and
\citet{blanc15}.

The so-called ``classical'' way to estimate oxygen abundances is the
electron temperature ($T_e$) method, which estimates the electron
temperature and density of a nebula using a number of oxygen lines in
different ionization states, including the auroral
[O~III]~$\lambda$4363 line, to then directly estimate the O~II and
O~III abundances and finally, after correcting for the unseen stages
of ionization, to obtain the total oxygen abundance.
However, except for low-metallicity environments, the auroral
[O~III]~$\lambda$4363 line is very weak, and it saturates at
metallicities higher than solar \citep{stasinska02}, since at higher
metallicities the cooling is dominated by the oxygen fine structure
lines in the near-infrared (NIR).  In addition, there are other
caveats about the $T_e$-method, as fluctuations and gradients in
temperature or in chemical composition may lead to underestimates of
the oxygen abundance (see for example \citealt{peimbert67}, and
\citealt{lopezsanchez12}).  Most recently, \citet{berg15} suggested
that among the auroral lines [O~II], [O~III] and [N~II], the [O~III]
$\lambda$4363 line, commonly used for $T_e$ measurements, is the most
problematic one, giving rise to temperature discrepancies.  Thus,
other methods have been developed that estimate the oxygen abundance
from ratios of strong nebular lines in the spectra of HII regions,
including amongst others [O~III] $\lambda \lambda$ 4959, 5007, [O~II]
$\lambda \lambda$ 3726, 3729, [N~II] $\lambda$ 6584, [S~II] $\lambda
\lambda$ 6717, 6731, as well as H$\alpha$ and H$\beta$. These are
called \emph{strong-line calibrations} and are the subject of this
manuscript.\footnote{There is one more class of methods, for which the
  recombination lines of different metal ions are used (e.g.,
  \citealt{stasinska02,lopezsanchez12}). However, these lines are so
  weak (for O and C they are $\sim10^{-4}$ of H$\beta$) that these
  methods can be used for HII regions in the Milky Way and in the
  Local Group, but not over large extragalactic distances, in which we
  are interested.}

Strong-line methods can be categorized into three types, depending on
how they calibrate the observed emission line ratios:
\begin{itemize}
\item{ empirical methods, which calibrate line ratios on observed
  $T_e$-based metallicities (e.g., \citealt{pilyugin05} -- hereafter
  P05; \citealt{pilyugin10} -- P10; \citealt{pilyugin12} -- P12;
  \citealt{marino13} -- M13), }
\item{ theoretical methods, which rely on calibrating various observed
  line ratios using stellar population and photoionization models
  (basically theoretically simulating HII regions, e.g.,
  \citealt{mcgaugh91} -- hereafter M91, KD02,
  \citealt{tremonti04})}
\item{ and lastly hybrid methods that use a mixture of both empirical
  and theoretical calibrations (e.g., \citealt{denicolo02} -- hereafter
  D02, \citealt{pettini04} -- PP04; \citealt{maiolino08} -- M08; and
  \citealt{perezmontero14} -- PM14).  }
\end{itemize}

The ionization parameter $q$ needs to be taken into account in all
calibrations, as HII regions at the same metallicity but with
different ionization parameters produce different line strengths. The
importance of the ionization parameter, which can be physically
understood as corresponding to the maximum velocity of an ionized
front that can be driven by the local radiation field of hot massive
stars that are ionizing the ISM gas, was demonstrated in a number of
works including \citealt{evans85,dopita86}, KD02, PM14, and
\citealt{sanchez15}. All calibrations include its impact, either
implicitly (by assuming the input object has the same values as the
calibration sample) or by explicitly solving for it.

While historically there have been large systematic offsets between
the $T_e$ method and the strong line methods, partially due to real
temperature gradients that exist in the high-ionization zones of the
HII regions \citep{lopezsanchez12}, D13 demonstrated that the results
of the $T_e$ method and strong line methods are reconciled if the
energy distribution of the electrons in the HII regions is not assumed
to be a simple Maxwell-Boltzmann distribution (as in prior works), but
a more realistic $\kappa$ distribution \citep{vasyliunas68, owoki83},
as observed in Solar plasma. In addition they find that the effect of
changing the specific $\kappa$ distribution on the strong-line methods
is minor (see also \citealt{Mendoza14}).

There is a long-standing debate about which diagnostic to use, and
there are systematic metallicity offsets between different methods
(recombination lines vs. strong-line method vs. the ``direct'' $T_e$
method), however, \emph{the relative metallicity trends can be
  considered robust, if the analysis is performed self-consistently in
  the same calibrator, and trends are seen across different
  calibrators} (KD08, \citealt{moustakas10}). It is of course
necessary to estimate uncertainties for the relative comparisons to be
meaningful. Note that while conversion values between different
calibrators \citep{kewley08} have been estimated, they are average
quantities derived from large data sets (tens of thousands of SDSS
galaxies), and they should be used with caution, if at all, on
individual metallicity measurements.

\section{Commonly Used metallicity indicators}
Line ratios commonly used as metallicity indicators (used in PP04,
P05, P12, and M13) are $$N2~=~\frac{\mathrm{[N~II]}\lambda6584}{\ha}$$
and the $O3N2$ index, first introduced by \citet{alloin79} and defined
as $O3/\hb/N2$ where $O3/\hb~=~
\frac{\mathrm{[O~III]}\lambda4959+\mathrm{[O~III]}\lambda5007}{\hb}$,
thus:
$$O3N2~=~\frac{\mathrm{[O~III]}\lambda4959+\mathrm{[O~III]}\lambda5007}{\hb}
\times \frac{\ha}{\mathrm{[N~II]}\lambda6584}.$$





Since $N2$ employs two closely spaced lines (\ha~ and
[N~II]$\lambda6584$), their ratio is not affected by stellar
absorption or (uncertain) reddening, and they are easily observed in
one simple spectroscopic setup. Thus the PP04 calibration based on
$N2$ has become a popular calibrator to use for low-$z$ SN host galaxy
studies (see meta-analyses by
\citealt{sanders12,modjaz12_proc,leloudas14} and others). However, it
is important to note that the PP04-N2 calibration has a number of
short-comings: the PP04 calibration implicitly assumes the same
relationship between the line ratios and the ionization parameter as
in the calibration sample from which it was initially derived, which
includes only 137 extragalactic HII regions, and it is a hybrid
calibration method. An updated calibration by M13 is based on a sample
of HII regions almost three times larger than that of PP04 and, more
importantly, \emph{only} uses $T_e$-based metallicities. M13 finds a
significantly shallower slope in the relationship between the $N2$ and
$O3N2$ indices and the oxygen abundance. Our code outputs M13 as a
default, and PP04 if explicitly requested.  It is important to note
that the employed nitrogen emission line saturates at high
metallicity, and thus the $N2$ methods, including M13, fail at
high-metallicity galaxies -- at $\oxabinline > 8.8$ (KD08).

Additional strong line indicators include the $N2O2$ ratio:
$$N2O2 ~=~ \frac{[N~II]~\lambda 6584}{[O~II]~\lambda 3727},$$
introduced by KD02, which exploits the leverage offered by the large
diatance in wavelength of the $[N~II]~\lambda 6584$ and
$[O~II]~\lambda3727$ lines, but is then very sensitive to reddening
and thus requires the measurements of $\ha$ and $\hb$ to be available,
and the Oxygen based $O3O2$ indicator:
$$O3O2=\frac{\mathrm{[O~III]}\lambda4959+\mathrm{[O~III]}\lambda5007}{\mathrm{[O~II]}\lambda3727},$$
already introduced by M91, which has the obvious advantage of using
only Oxygen lines, which are, as mentioned, the strongest emission
lines in optical wavelengths apart from Hydrogen.

The ratio of oxygen line fluxes to \hb~is referred to as $R_{23}$
\citep{pagel79}:
$$R_{23}=\frac{\mathrm{[O~II]} \lambda 3727~+~\mathrm{[O~III]}
  \lambda\lambda 4959, 5007}{\hb},$$ where [O~III]~$\lambda\lambda$
4959,5007 stands for the sum of the two [O~III] lines.  This popular
indicator is double-valued with metallicity: the same $R_{23}$ value
may correspond to two metallicity values, and thus other indicators
need to be used to break the degeneracy between the high (``upper
branch'') and the low values (``lower branch'') of the $R_{23}$
metallicities (e.g., KD08, \citealt{moustakas10}). A similarly double
valued indicator is $S_{23}$, which is based on the sulfure lines: $
S_{23}=( \mathrm{[S~II]}~\lambda6717+\mathrm{[S~III]}~\lambda9069) /
\hb $.  For these double-values indicators the curvature of the
relationship between the line ratio and the metallicity in the two
branches depends on the ionization parameter, and both empirical and
theoretical calibrators that use them have to solve for $q$.  For the
theoretical strong-line method, calibrations of $R_{23}$ by M91, KD02,
and D13 use different theoretical photoionization models: different
stellar populations codes (e.g., Starburst 99, based on
\citealt{bruzual93}, or POPSTAR, \citealt{molla09}) and
photoionization codes (e.g., MAPPINGS \citealt{allen08}, or CLOUDY,
\citealt{ferland98}) to calibrate different line ratio indices.  M91
and KD02 use an iterative process to break the $R_{23}$ degeneracy and
to constrain the ionization parameter $q$. Other calibrations, such as
Z94, do not explicitly solve for $q$.  On the empirical calibrators
side, P10 introduces the $P$ parameter, which is a modified version of
the ionization parameter $q$.

As can be seen, each calibrator has different advantages and
disadvantages and should be used in different metallicity regimes (see
detailed discussion in, for example, KD02, \citealt{stasinska02},
KD08, \citealt{moustakas10,lopezsanchez12}, D13,
\citealt{blanc15}). Thus, \verb=pyMCZ= outputs the oxygen abundance in
the main metallicity calibrators whenever the lines necessary to
calculate the indicators used by each method are available. As default
output values for 11 metallicity diagnostics are generated (version
v1.3, Fall 2015). By ``diagnostics'' here we mean a suite metallicity
calibrators described in a single paper (e.g. KD02), and some of these
diagnostics describe multiple calibrators based on different
indicators: the KD02 diagnostic have four outputs and the PP04, P10,
M13 , M08, and D13 diagnostic have multiple outputs as well.

\section{pyMCZ: a python module for strong-line metallicity diagnostics}\label{sec:method_sec}

\begin{figure*}[!ht]
\centerline{
  \includegraphics[width=0.98\columnwidth]{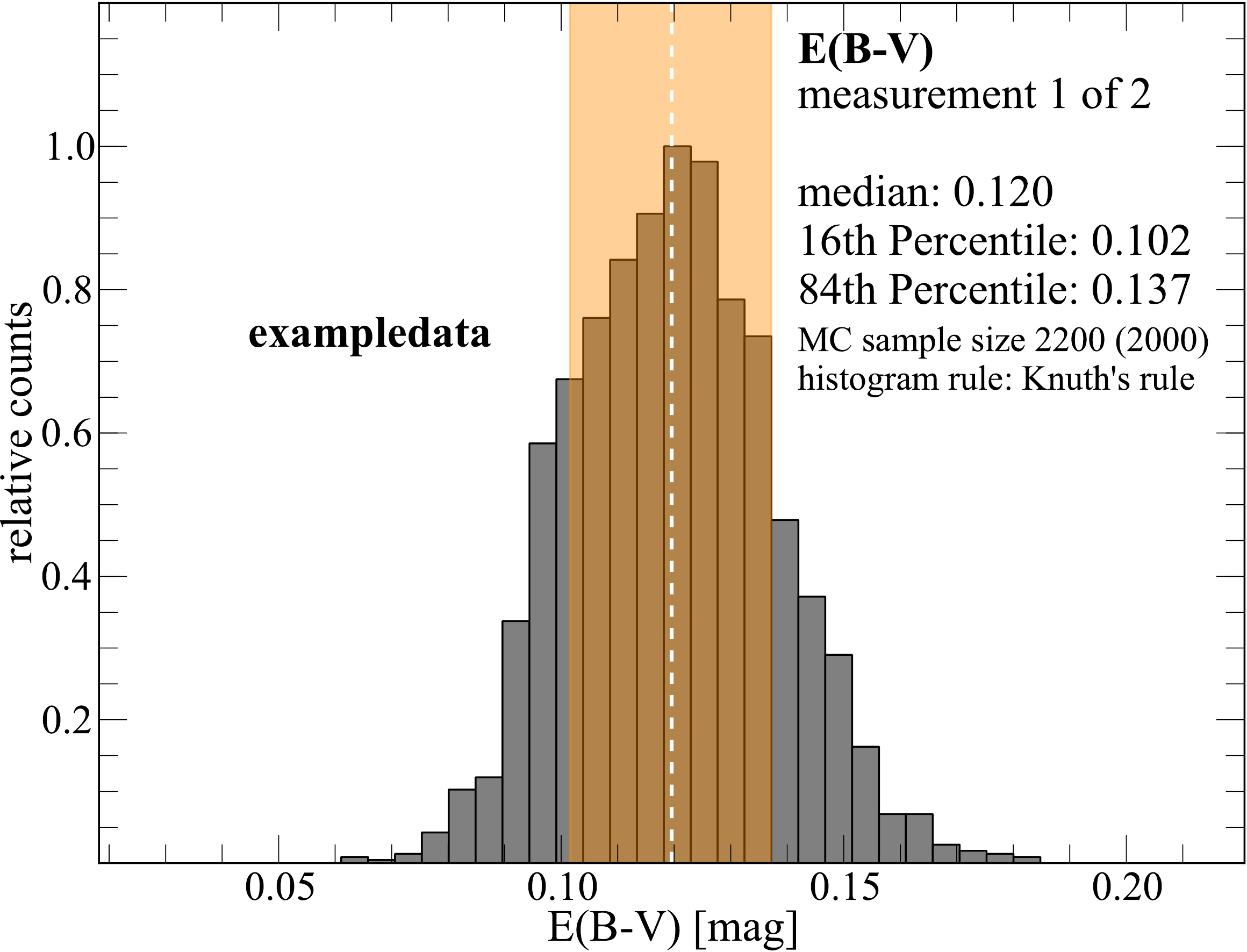}
  \includegraphics[width=0.98\columnwidth]{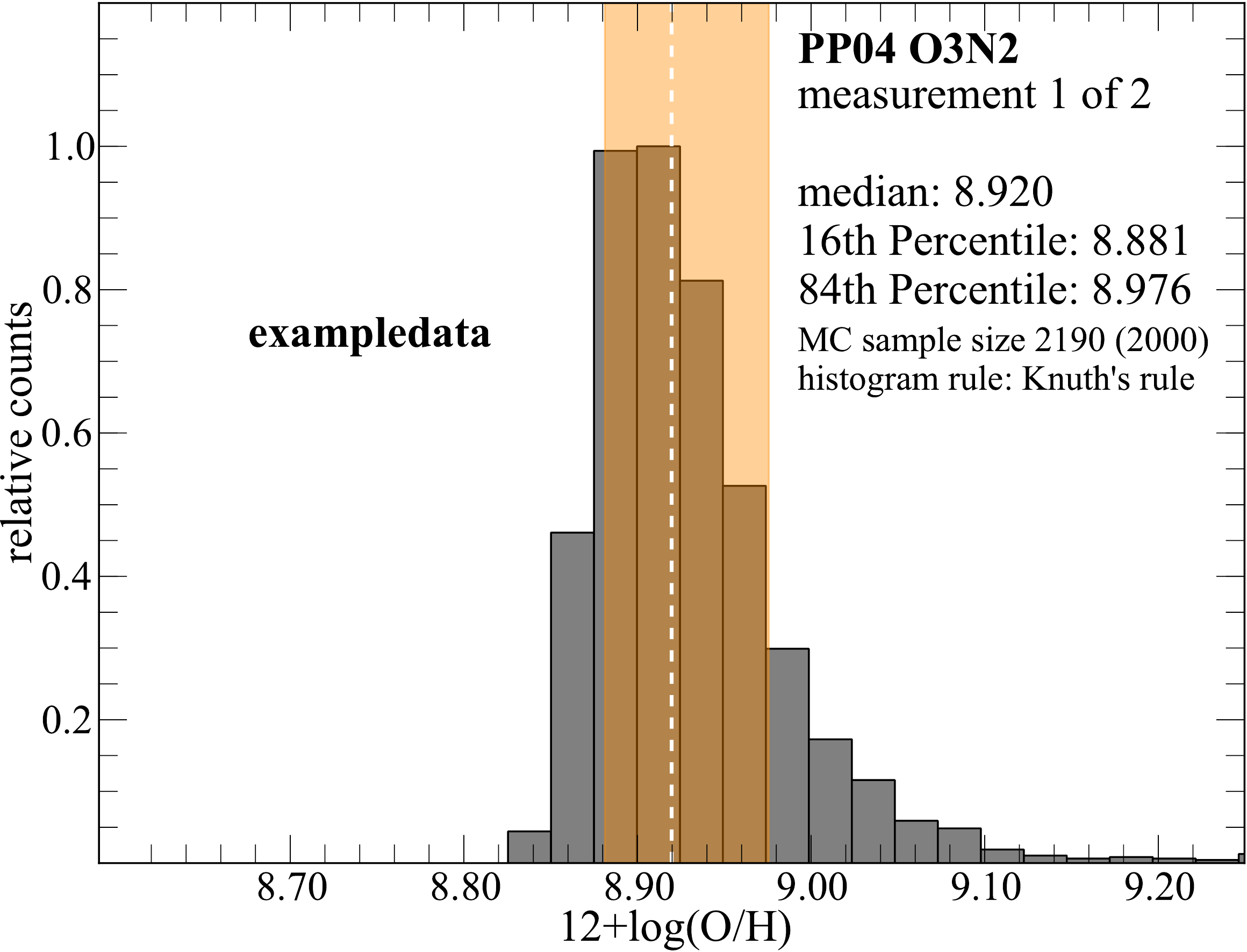}}
\centerline{
  \includegraphics[width=0.98\columnwidth]{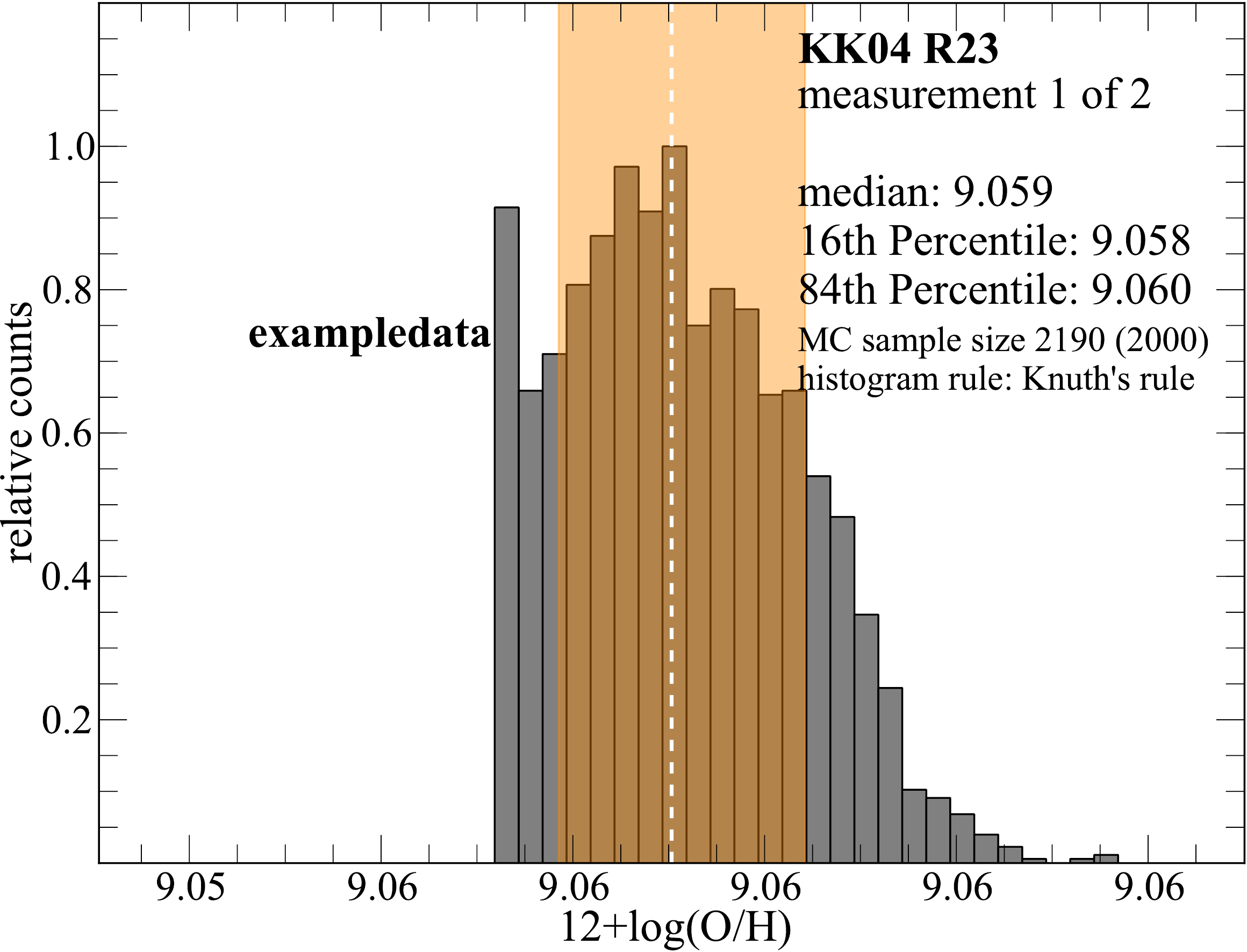}
  \includegraphics[width=0.98\columnwidth]{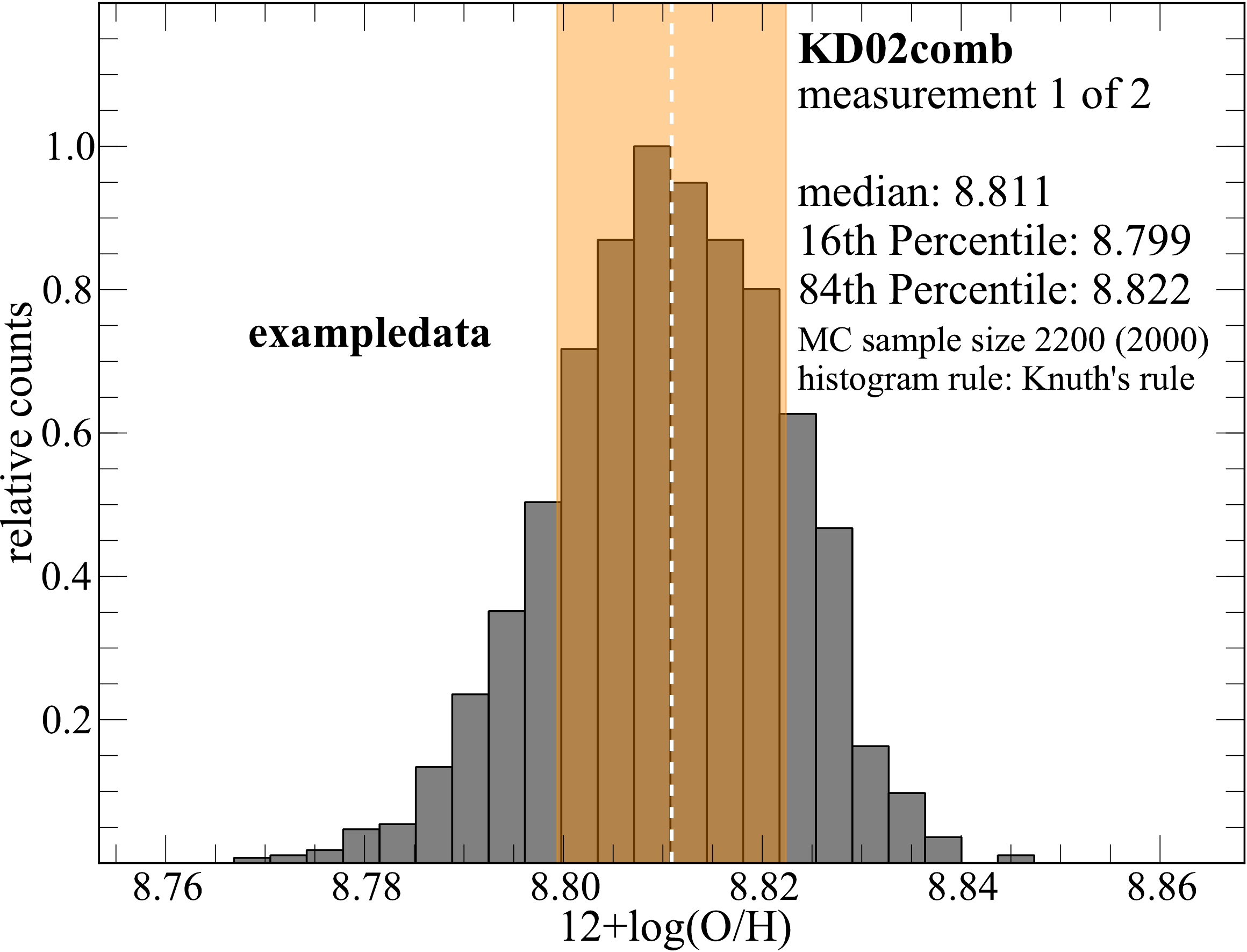}}
\caption{Metallicity and reddening \ebmv~ parameter distributions
  based on the example data shown in \autoref{tab:exampledata}:
  emission line data of the HII regions at the position of SN~2008D,
  published in \citealt{modjaz11} (see \autoref{sec:comp_sec} for a
  discussion of the test sample). The distributions are generated from
  $N$=2,000 samples. The median values are shown with dashed white
  lines, while the confidence region, between the 16$^{th}$ and the
  84$^{th}$ percentile is shaded in orange. We show the metallicity
  calibrators from \citet{pettini04}, using [O~III] and [N~II]
  (PP04\_O3N2), the $R_{23}$-based calibrator described in
  \citet{kobulnicky04} (KK04\_$R_{23}$), and the combined calibrator
  of \citet{kewley02} (KD02\_comb), updated as described in
  \citet{kewley08}. Similar plots are outputted by the code for all
  computed metallicity calibrators. We note PP04 calibrators are an
  \emph{optional} outputs of our code, since they are superseded by
  M13 (see \autoref{sec:diags}). However, we use them for
  comparison with the results obtained in \citealt{modjaz11}. Each
  plot indicates the calibrator, the sequential number of measurement
  in the input file (here ``measurement 1 of 2'', which corresponds to
  the first line of the input ASCII files), the median, 16$^{th}$ and
  84$^{th}$ percentile values, the sample size (which is by default
  initially set to 10\% larger than the requested $N$ value, but can
  be smaller if some simulations lead to invalid metallicities), and,
  finally, the method used to choose the bin size for the histogram
  (Kuth's rule in this case, see \autoref{sec:vizs}).} 
\label{fig:metallicity_distribution}
\end{figure*}

We developed a Python package, \verb=pyMCZ= for the simultaneous
calculation of \oxabinline~ in various calibrators, and its confidence
region.  We started with the iterative IDL code by KD02 hereafter
referred to as IDLKD02, specifically a version updated in KE08. We
translated the code into Python and added new, more recent calibrators
(see \autoref{sec:IO}) and new features, of which the most important
is the capability of obtaining uncertainties on the metallicity
outputs via MC sampling.

\subsection{Availability and Technical details}
The source code is published under
MIT-license\footnote{\url{https://github.com/nyusngroup/pyMCZ/blob/master/LICENSE.txt}}
on GitHub.\footnote{\url{https://github.com/nyusngroup/pyMCZ}} At this
time the code is released under DO~I: 10.5281/zenodo.17880 and the
most recent stable version is version 1.3: {\b pyMCZ v1.3}. Project
details, step-wise tutorials, and further information can be found in
the module
README.\footnote{\url{https://github.com/nyusngroup/pyMCZ/blob/master/README.md}}
Development is done in Linux and OS X. The package requires standard
python packages, such as \verb=numpy=, \verb=scipy=, \verb=pylab=, and
additional features are enabled if the packages \verb=astroml=,
\verb=skitlearn=, and \verb=pyqz= are installed, but these packages
are \emph{not} required.  \verb=pyMCZ= can be
run as a standalone package from the local directory, or it can be
installed (via \verb=python setup.py install=).

\subsection{Input and Output of code}\label{sec:IO}

\begin{deluxetable*}{lccc} 
\tabletypesize{\normalsize} \setlength{\tabcolsep}{0.000000in}
\tablecolumns{10} \tablecaption{calibrators and line ratios/indicators
  from which the calibrators are derived} \tablehead{ 
  \colhead{Calibrator \tablenotemark{a}} & \colhead{Lines used for
    metallicity calculation \tablenotemark{b}} & \colhead{Support
    lines and calibrators \tablenotemark{c}} & \colhead{Reference} }
\startdata {\footnotesize M91} & $R_{23}$, $O3O2$ & $N2$, $N2O2$ &
\citet{mcgaugh91}\\ {\footnotesize Z94*} & $R_{23}$ & - &
\citet{zaritsky94}\\ {\footnotesize (DP00*)} & $S_{23}$ & - &
\citet{diaz00}\\ {\footnotesize (P01)} & $R_{23}$, $O3/\hb$,
$\mathrm{[O~II]}\lambda3727/\hb$ & $N2$, $N2O2$ &
\citet{pilyugin01}\\ \\{\footnotesize (C01\_R23)} & $R_{23}$,
${\mathrm{[O~III]}\lambda5007}/{\hb}$,
      {\normalsize$\frac{\mathrm{[O~II]}\lambda3727}{\mathrm{[O~III]}\lambda5007}$}
      & - & \citet{charlot01}\\ \\{\footnotesize (C01\_N2S2)} &
      $R_{23}$,
      {\normalsize$\frac{\mathrm{[O~II]}\lambda3727}{\mathrm{[O~III]}\lambda5007}$},
      {\normalsize$\frac{\mathrm{[N~II]}\lambda6584}{\mathrm{[S~II]}\lambda6717}$}
      & - & \citet{charlot01}\\ \\{\footnotesize D02*} & $N2$ & - &
      \citet{denicolo02}\\ {\footnotesize KD02\_N2O2*} & $N2O2$ & - &
      \citet{kewley02}\\ {\footnotesize (PP04\_N2Ha*)} & $N2$ & - &
      \citet{pettini04} \\ {\footnotesize (PP04\_O3N2*)} & $N2$,
      $O3/\hb$ & - & \citet{pettini04} \\ {\footnotesize KK04\_N2Ha} &
      $N2$, $q$ & $N2O2$ &\citet{kobulnicky04}\\ {\footnotesize P05} &
      $R_{23}$, $O3/\hb$, ${\mathrm{[O~II]}\lambda3727}/{\hb}$ & $N2$,
      $N2O2$ &\citet{pilyugin05}\\ {\footnotesize KK04\_$R_{23}$} &
      $R_{23}$, $q$, & $N2$, $N2O2$ &
      \citet{kewley08}\\ {\footnotesize KD02\_comb} \tablenotemark{d}&
      M91, KD02\_N2O2, KD02\_N2Ha, K{\footnotesize D04\_R23 }& $N2$,
      $N2O2$ &\citet{kewley08}\\ \\{\footnotesize M08\_O3O2 }& $O3O2$
      & - & \citet{maiolino08}\\ \\{\footnotesize M08\_N2Ha }& $N2$ &
      - & \citet{maiolino08}\\ {\footnotesize M08\_R23 }& $R_{23}$ &
      M08\_O3O2, M08\_N2Ha & \citet{maiolino08}\\ {\footnotesize
        (M08\_O3Hb) }& ${\mathrm{[O~III]}\lambda5007}/{\hb}$ &
      M08\_O3O2, M08\_N2Ha & \citet{maiolino08}\\ {\footnotesize
        (M08\_O2Hb) }& ${\mathrm{[O~II]}\lambda3727}/{\hb}$ &
      M08\_O3O2, M08\_N2Ha & \citet{maiolino08}\\ \\{\footnotesize
        (M08\_O3N2) }&
      {\normalsize$\frac{\mathrm{[O~III]}\lambda5007}{\mathrm{[N~II]}\lambda6584}$}
      & M08\_O3O2, M08\_N2Ha & \citet{maiolino08}\\ \\{\footnotesize
        P10\_ON }& $N2$, $O3/\hb$,
      ${\mathrm{[O~II]}\lambda3727}/{\hb}$ & - &
      \citet{pilyugin10}\\ {\footnotesize P10\_ONS }&
      ${\mathrm{[N~II]}\lambda6584}/{\hb}$, $O3/\hb$& - &
      \citet{pilyugin10}\\ \\&${\mathrm{[O~II]}\lambda3727}/{\hb}$,
            {\normalsize$\frac{\mathrm{[S~II]}\lambda6717+\mathrm{[S~II]}\lambda6731}{\hb}$}
            & &\\ \\{\footnotesize M13\_N2* }&
            ${\mathrm{[N~II]}\lambda6584}/{\hb}$ & -
            &\citet{marino13}\\ {\footnotesize M13\_O3N2* }&
            ${\mathrm{[N~II]}\lambda6584}/{\hb}$, $O3/\hb$ & -
            &\citet{marino13}\\ \\{\footnotesize D13\_N2S2\_O3S2 }&
            {\normalsize$\frac{\mathrm{[N~II]}\lambda6584}{\mathrm{[S~II]}\lambda6717+\mathrm{[S~II]}\lambda6731}$},
            {\normalsize$\frac{\mathrm{[O~III]}\lambda5007}{\mathrm{[S~II]}\lambda6717+\mathrm{[S~II]}\lambda6731}$}
            & - &\citet{dopita13}\\ \\{\footnotesize D13\_N2S2\_O3Hb
            }&
            {\normalsize$\frac{\mathrm{[N~II]}\lambda6584}{\mathrm{[S~II]}\lambda6717+\mathrm{[S~II]}\lambda6731}$},
            $O3/\hb$ & - &\citet{dopita13}\\ \\{\footnotesize
              D13\_N2S2\_O3O2 }&
            {\normalsize$\frac{\mathrm{[N~II]}\lambda6584}{\mathrm{[S~II]}\lambda6717+\mathrm{[S~II]}\lambda6731}$},
            $O3O2$ & - &\citet{dopita13}\\ \\{\footnotesize
              D13\_N2O2\_O3S2 }& $N2O2$,
            {\normalsize$\frac{\mathrm{[O~III]}\lambda5007}{\mathrm{[S~II]}\lambda6717+\mathrm{[S~II]}\lambda6731}$}
            & - &\citet{dopita13}\\ \\{\footnotesize D13\_N2O2\_O3Hb
            }& $N2O2$, $O3/\hb$ & -
            &\citet{dopita13}\\ \\{\footnotesize D13\_N2O2\_O3O2 }&
            $N2O2$, $O3O2$ & - &\citet{dopita13}\\ \\{\footnotesize
              D13\_N2Ha\_O3Hb }& $N2$, $O3/\hb$ & -
            &\citet{dopita13}\\ \\{\footnotesize D13\_N2Ha\_O3O2 }&
            $N2$, $O3O2$ & - &\citet{dopita13}\\ \enddata
            \tablenotetext{a}{Calibrators in parentheses are optional
              outputs of our pyMCZ.}  \tablenotetext{b}{The flux
              ratios used to calculate the metallicity.}
            \tablenotetext{c}{Additional flux ratios used to estimate
              the metallicity regime, select the upper or lower branch
              in double valued metallicity solutions, initiate
              minimization, etc.}  \tablenotetext{d}{This combined
              calibrator uses the output of other metallicity
              calibrators in different regimes and outputs a
              ``recommended'' value appropriate for that abundance
              regime.}  \tablenotetext{*}{These calibrators use fully
              linear calculations and the observational uncertainty
              could therefore be propagated analytically.}
\label{tab:calibrators}
\end{deluxetable*}

The input of the code is a set of spectral emission line fluxes. We
assume that the observed emission lines originate in HII regions and
are not due to non-thermal excitation by, for example, AGN,
interstellar shocks from SNe, or stellar winds.  Tests to exclude data
contaminated by such non-thermal sources should be executed
\emph{prior to running this code} using the recommended indicators by,
e.g., \citet{baldwin81}, \citet{kauffmann03}, \citet{kewley06_sdss},
and \citet{cidfernandes10,sanchez15}. Furthermore, these lines should
have all the correct flux calibration (at least correct relative
calibration) and \emph{should have a SNR of at
  least 3}.
The latter requirement is important for the success of the sampling
technique. In our code, synthetic line flux measurements are drawn
within a Gaussian distribution with standard deviation equal to the
measurement error, and centered on the measured flux value
(\autoref{sec:uncert}). Thus, a $\mathrm{SNR} >= 3$ assures that
fewer than $\sim1\%$ of the sampled fluxes fall below zero (and are
thus invalid).  The code checks if for any line the $\mathrm{SNR} >=
3$ condition is not satisfied, and issues a warning message.

Emission line flux values are fed into our Python implementation as in
the original IDLKD02 code. The inputs are emission line flux values,
and their uncertainties, for the following lines: \ha, \hb,
[O~I]~$\lambda$ 6300, [O~II]~$\lambda$ 3727, [O~III]~$\lambda$ 4959,
[O~III]~$\lambda$ 5007, [N~II]~$\lambda$ 6584, [S~II]~$\lambda$ 6717,
[S~II]~$\lambda$ 6731, [S~III]~$\lambda$ 9096 and [S~III]~$\lambda$
9532. The latter two lines can be used to calculate the sulfur ratio
$S_{23}$, 
but are not often observed since they are in the NIR. Only one
metallicity calibrator based on $S_{23}$, \citealt{diaz00} (DP00), is
implemented in the current version of the code.  The line fluxes are
to be stored in an ASCII file, and the measurement errors in a
separate ASCII file (for consistency with the original IDLKS02 code;
consult the
README.md\footnote{\url{https://github.com/nyusngroup/pyMCZ/blob/master/README.md}}
in the GitHub repository for details about the input format, and find
example files in the repository).  If the fluxes for the specified
lines are not available, the entry should be set to \emph{NaN}. The
oxygen abundance will be calculated only for metallicity calibrators
that use valid, non-\emph{NaN}, line fluxes.  If the line fluxes
necessary for specific calibrators are not provided, the output
metallicities will default to \emph{NaN}. In absence of measurement
uncertainties, the error should be set to 0.0 in the input ASCII
file. If the errors in the measurements are not provided, the code
will specify that it cannot create a measurement distribution or
determine a confidence interval, but it will calculate and output the
nominal metallicity obtained from the input flux measurement.

The inputted line fluxes are corrected for reddening by using the
observed Balmer decrement, for which \ha~ and \hb~ flux values need to
be provided. We assume case B recombination, and thus the standard
value of 2.86 as the intrinsic \ha/\hb~ ratio \citep{osterbrock89},
and apply the standard Galactic reddening law with $R_V$ = 3.1
\citep{cardelli89}. However, the user can choose other extinction laws
and $R_V$ values, if desired, given the code's open-source nature. If
the input measurements are already de-reddened, the user can disable
the reddening correction. If either \ha~ or \hb~ are not provided, the
reddening correction cannot be implemented. The user is notified with
a warning message and has the option to proceed with the calculations
with uncorrected line fluxes. This option is enabled, and should of
course only be chosen, for cases in which the spectral fluxes are
already dereddened before being fed is input to \verb=pyMCZ=.

We implement the calculation of \oxabinline~ values and their
uncertainties from the metallicity calibrators listed below, and
summarized in \autoref{tab:calibrators}, implemented as prescribed
in KE08 where not otherwise noted. We emphasize that subsets of these
calibrators use the same input line ratios, and that related
assumptions and similar observation and simulations are used to
generate the calibrator algorithms (e.g. there are several calibrators
that are essential linear or near-linear functions of $N2$: PP04\_N2,
KK04\_N2 and M13\_N2). Thus the oxygen abundance results produced
using different calibrator are far from independent.

\begin{itemize}
\item {\bf M91} \citep{mcgaugh91}
 is an $R_{23}$-based theoretical calibrator which
  also determines the ionization parameter. To break the $R_{23}$ degeneracy we use  $N2O2$, following KE08 Appendix 2. 
\item {\bf Z94} \citep{zaritsky94} which is valid for the upper branch
  of $R_{23}$ only, and we conservatively constrain it to
  log($R_{23})<0.9$, the range that is covered by the
  photoionization model grids.
\item {\bf D02} \citep{denicolo02} for which we include, in addition
  to the uncertainties in the measurements, the uncertainty in the fit
  parameters as they appeared in the original paper.
\item {\bf P05} \citep{pilyugin05} an $R_{23}$-based method calibrated via $T_e$ metallicities on a sample of HII regions. We use the values of $N2O2$, and $N2$ to discriminate between  upper and lower branch.
\item {\bf KD02 \& KK04}: 4 calibrators: KD02\_N2O2, which uses the
  $N2O2$ indicator~
\citep{kewley02}, KK04\_N2\ha~ which uses the N2
indicator~\citep{kobulnicky04}, KK04\_$R_{23}$ (\citealt{kewley08},
appendix A2.2), which is based on the value of $R_{23}$, and a
combined method, KD02\_comb that chooses the optimal method given the
input line fluxes and is implemented as described in Appendix 2.3 of
\citet{kewley08}.
\item {\bf M08} \citep{maiolino08}: This calibrator is a combination
  of the KD02 photoionization models at high metallicities and $T_e$
  based metallicities at low metallicities. Our default output
  includes their strong line diagnostic, which is based on $R_{23}$, and
  the diagnostics based on $O3O2$ and [N~II]/\ha, since the
  metallicity estimates from [N~II]/\ha~ or $O3O2$ are necessary to
  resolve the degeneracy in the double-valued $R_{23}$
  metallicity. The other calibrators (based on [O~II]/\hb,
  [O~III]/\hb, and [O~III]/[N~II]) can be outputted upon explicit user
  request, via the command line options.
\item {\bf P10} \citep{pilyugin10}: This is the so-called ONS
  diagnostic (involving the [O~II], [O~III], [N~II] and [S~II] lines)
  and is calibrated with HII regions that have $T_e$ based
  metallicities.
\item {\bf M13} \citep{marino13}: Two calibrators: M13\_N2, which is a
  linear fit to the $\mathrm{[N~II]}~\lambda6584/\ha$, and
  M13\_O3N2. This is an updated calibration of the PP04 $O3N2$ method,
  based on a large number of $T_e$-based metallicity measurements,
  including those from the Calar Alto Legacy Integral Field
  spectroscopy Area survey (CALIFA, \citealt{sanchez12}, almost three
  times larger than the sample used in PP04). This method derives a
  significantly shallower slope between the $O3N2$ index and the
  oxygen abundance than the PP04 calibration did.
\item {\bf D13}: \citep{dopita13}. The photoionization models used in
  KD02 and in KK04 are updated by including new atomic data within a
  modified photoionization code and by assuming a $\kappa$
  distribution for the energy of the electrons in the HII regions,
  rather than the simple Maxwell-Boltzmann distribution assumed in
  prior works. This distribution is more realistic, as observed in
  Solar plasma \citep{nicholls12}. If the user has installed the
  publicly available
  \verb=pyqz=\footnote{\url{https://datacommons.anu.edu.au:8443/DataCommons/item/anudc:5037}}
  Python module, [N~II], [S~II], [O~III], \ha, and \hb~ lines are fed
  to the \verb=pyqz= module, which produces up to eight emission line
  ratio diagnostics for \oxabinline, each using two of the indicators
  [N~II]/[S~II], [N~II]/\ha, [O~III]/[S~II], and [O~III]/\hb. Our code
  sets the $\kappa$ parameter to 20, which is the value that D13 found
  best resolves the inconsistencies between oxygen abundance values
  derived from the strong-line methods and from the ``direct'' $T_e$
  method\footnote{The user can modify the value of $\kappa$ by editing
    the code, if they wish. The user may then need to generate the
    appropriate diagnostic grids, see \texttt{pyqz} ’s instructions
    \url{http://fpavogt.github.io/pyqz/} for further details. The
    current version of \texttt{pyMCZ} is compatible with versions 0.5.0 through
    0.7.2 (current version at the time of publishing this work) of  \texttt{pyqz}.}.

\item{\bf DP00, P01, C01 \& PP04 (only upon request)}:
  (\citealt{diaz00}, \citealt{pilyugin01}, \citealt{charlot01}). DP00
  is based on $S_{23}$, and is the only $S_{23}$ calibrator
  implemented in the current version of the code. P01 is superseded by
  P05. C01 produces a diagnostic based on $R_{23}$, C01\_$R_{23}$, and
  one based on [N~II]/[S~II], C01\_N2S2; C01\_$R_{23}$ was used to
  calculate the combined calibrator described in KK04, and included in
  the IDLKD02 code, but it is no longer used in the new combined
  calibrator KD02\_comb, which supersedes the old one. Thus, P01 and
  C01 are only included for historical reasons.  PP04:
  \citep{pettini04} includes two calibrators PP04\_N2, based on the
        [N~II]/\ha~ ratio, and PP04\_O3N2, based on
        ($\frac{\mathrm{[O~III]}}{\hb}/\frac{\mathrm{[N~II]}}{\ha}$).
        calibrators are commonly used in the literature investigating
        SN and GRB environments (see \citealt{modjaz11} and referenced
        therein). They are superseded by the M13 calibrators (see
        \autoref{sec:diags}), but are included here for
        completeness.  These five calibrators are \emph{not} part of
        the default output of our code, however they are still
        available upon explicit user request via command line input.

\end{itemize}

The following diagnostics were calculated in the original IDLKD02
code, and are discussed in detail in KD02 and \citet{kewley08}: C01,
P01, M91, Z94, D02, PP04, P05, KD02, KK04, and KD02\_comb. We refer
the reader to those papers for further details. DP00 is the only
diagnostic that relies on sulfur ratios $S_{23}$.  The shortcomings of
$S_{23}$ as a tool to measure abundances are discussed in KD02.
The DP00 diagnostic we implement is modified with the addition of a
term $\propto (c+S_{23}^3)^{-1}$ as suggested by KD02, which corrects
the tendency of the calibrator to systematically underestimate the
abundance, with a discrepancy growing larger at higher
metallicity. However, we point out that the scatter in metallicity
derived from this diagnostic compared to others remains high.

The distributions of \ebmv~and $\log(R_{23})$ values are also part of
the default output. Certain parameters, such as the ionization
parameter $q$ and the electron density (using the [S~II] lines) are
computed as long as the necessary lines are provided, but are not
outputted in the current version of our code $-$ however, since the
code is open-source, the reader can easily modify the code to suit
their needs.

\subsection{Computing Statistical Uncertainties}\label{sec:uncert}

Three kinds of uncertainties must be taken into account when
calculating metallicities: 1) the propagation of measurements errors
of the line fluxes; 2) the propagation of uncertainty in the
calibration (either from model-fitting or propagating the
uncertainties of the coefficients in the linear calibrations); 3) the
systematic offsets/errors between different calibrations with respect
to the ``true'' value of the metallicity.

In our work we account for errors due to the emission line measurement
uncertainties, and, when provided by the authors, to the calibration
uncertainties, and propagate them into the derived metallicities. The
propagation of the observational errors cannot, in most cases, be
conducted analytically, since many metallicity calculations rely on,
for example, decision trees. Thus the MC environment is necessary in
most cases to properly assess the effect of the measurement
uncertainties, and, for consistency and to generate a simple, modular
code, we adopt it as the method for uncertainty calculation in all
cases.  We provide the full metallicity distributions, for
\textbf{all} calculated calibrators. The latter is important since, as
we show, metallicity distributions are usually not symmetric and can
be multimodal (for example for the $R_{23}$ based calibrators). In
addition, we provide the full metallicity parameter distribution to the
user. We note that we do not address the third source of uncertainty:
any systematic errors between different calibrators. This is the
subject, for example, of \citet{lopezsanchez12}.

For every set of input line measurements we introduce a MC sampling
method to obtain iterations via random sampling within the measurement
uncertainties, and thus we obtain a robust result for statistical
error estimation (e.g., \citealt{efron79,hastie09,andrae10}).  Given a
dataset with error bars from which certain parameters are estimated,
MC sampling generates synthetic data samples from a given
distribution. We draw synthetic data from a Gaussian distribution
centered on each measured line flux value, with standard deviation
corresponding to the measurement uncertainty. The implicit assumption
is made that the line flux error is Gaussian distributed in nature,
and that the errors are independent. However, the user who wishes to
provide their own probability distribution for the emission line
uncertainties can easily modify the code to include the desired
distribution.  For each metallicity calibrator, and for each of the
$N$ values chosen randomly within the relevant emission line
distributions, we run the calculations that compute the metallicity.
By generating synthetic data, this method effectively simulates
conducting multiple experiments when repeating observations is
impractical or impossible, as is the case of the emission line flux
data.

The sample size $N$ is set by the user, and one should expect an
appropriate value of $N$ to be a few 1000s, depending on the
metallicity calibrator chosen and measurement errors (for example:
$N=2,000$ is determined to be sufficient for our example data, as
shown in \autoref{sec:completeness}).

A distribution of parameter estimates for the oxygen abundance is
generated, from which the median metallicity and its confidence region
are calculated, and the results are binned and visualized in an
outputted histogram (see \autoref{sec:vizs}). This is done for each
calibrator the user chooses to calculate. The fiftieth (50\%)
percentile, i.e.  the median, is reported as the estimated metallicity
value, and the 16$^{th}$ and 84$^{th}$ percentiles of the metallicity
estimate distribution, as its confidence region. The user can also
choose to output the full metallicity distribution as
binary\footnote{using the \emph{pickle} Python module} or ASCII files.

This MC sampling approach takes into account the impact of the
uncertain reddening (due to the uncertainties in the measurement of
the \ha~and \hb~fluxes), when the option for de-reddened metallicities
is chosen. For each synthetic set of measurements, a new reddening
value is calculated based on the sampled \ha~and \hb~fluxes, and used
to compute the de-reddened metallicity value, so that the derived
distribution of metallicity values naturally takes into account the
uncertainty in reddening. The median value, and confidence intervals
for \ebmv, as well as a distribution histogram, are outputted together
with the metallicity calibrators (first panel in
\autoref{fig:metallicity_distribution}). If either \ha~or \hb~are not
provided, however, no reddening correction can be applied. The
computed metallicity will not be reddening-corrected and the
\ebmv~output will be set to zero.

\begin{figure}[ht!]
\includegraphics[width=1.0\columnwidth]{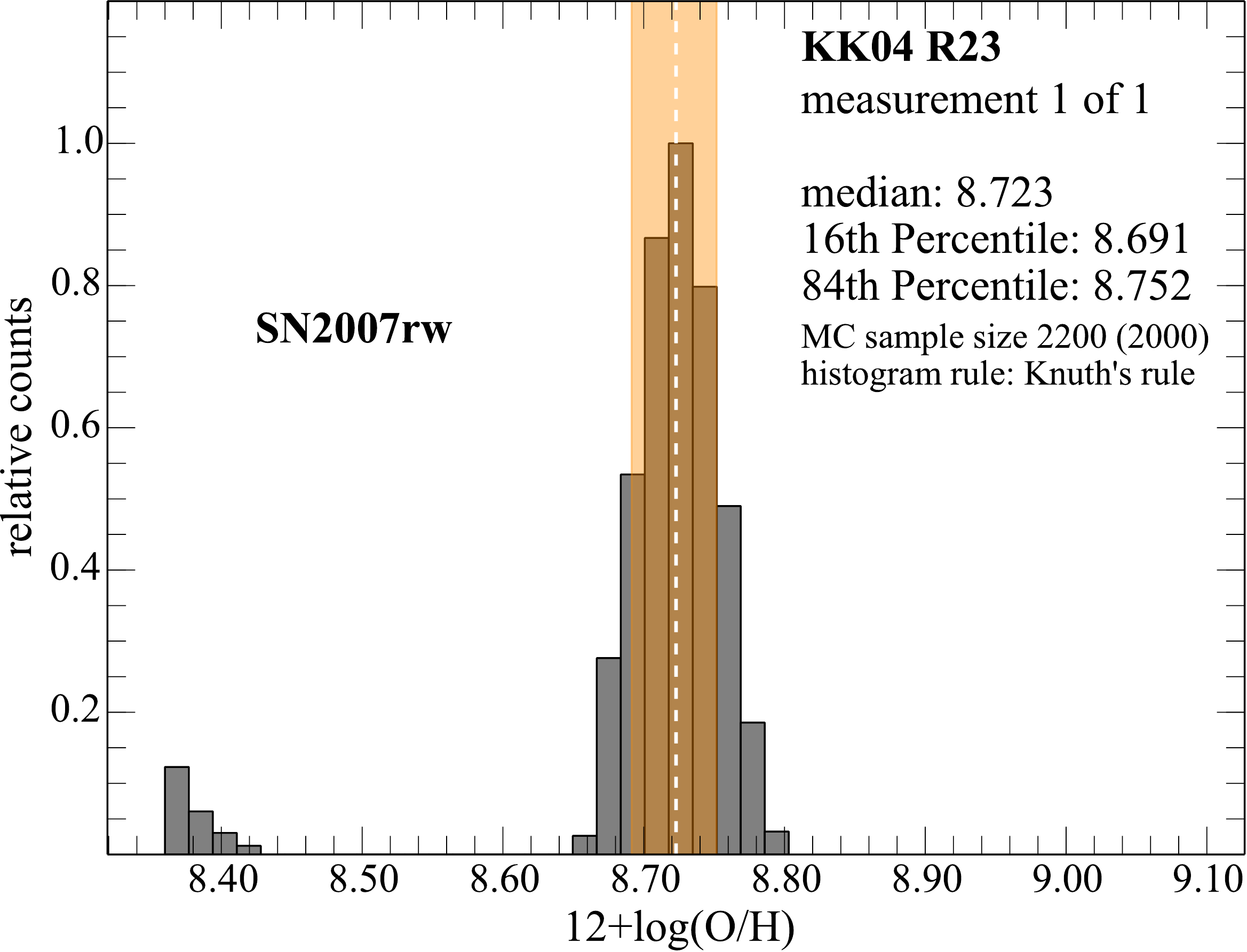}
\caption{Metallicity distribution according to the KK04\_$R_{23}$
  calibrator for the site of SN~2007rw, as measured in
  \citet{modjaz11}. The $R_{23}$ based calibrators are double valued:
  for the same $R_{23}$ value there are two metallicity solutions. The
  KK04\_$R_{23}$ calibrator uses the ionization parameter $q$ in an
  iterative fashion to determine the metallicity by selecting either
  the upper or lower branch value. In some cases, when the errors in
  the measurements are large, or for particular flux ratios, the
  solution may oscillate between the upper and lower branch in
  different realizations, giving rise to a bimodal metallicity
  distribution. These cases are easily identifiable by looking at the
  visual tools the code generates, such as this
  histogram.}\label{fig:metallicity_bimodal}
\end{figure}

\autoref{fig:metallicity_distribution} shows metallicity estimate
distributions for three representative calibrators (PP04\_O3N2,
KK04\_$R_{23}$, and KD02\_comb), and for the reddening parameter
\ebmv. Similar plots are produced for all calibrators calculated and
for $\log(R_{23})$ as a default output (and are available online in
the \verb=pyMCZ= GitHub
repository\footnote{\url{https://github.com/nyusngroup/pyMCZ}}). Although
the input line flux distributions are Gaussian, the metallicity
distributions are not, for two reasons: first, since the metallicities
are computed based on log values of line flux ratios, symmetric error
bars in linear space will translate into asymmetric error bars in log
space; and second, some metallicity calibrator computations are
non-linear, and sometimes bimodal, especially those that use $R_{23}$
and $S_{23}$, as shown in \autoref{fig:metallicity_bimodal}, since the
upper or lower branch metallicity value has to be chosen to break the
degeneracy for each synthetic measurement.

Since the metallicity distributions are not Gaussian, the percentiles
we report cannot be expressed in terms of $\sigma$ values. In
determining the confidence region for asymmetric and multi-modal
distributions, there are broadly three approaches
(\citealt{andrae10}): choosing a symmetric interval, the shortest
interval, or a \emph{central} interval.  With the central method, we
determine the confidence interval by choosing the left and right
boundaries such that the region outside the confidence interval on
each side contains $16\%$ of the total distribution $-$ in analogy to
the one-sigma-interval of a Gaussian distribution. This ensures that
the algorithm finds the proper boundaries even for asymmetric,
non-Gaussian distributions, and in the case of multiple peaks
(\autoref{fig:metallicity_bimodal}). In summary, the output for the
measured value corresponds to the fiftieth (50\%) percentile (median),
while the lower error bar corresponds to the 50$^{th}$-16$^{th}$
percentile and the upper error bar corresponds to 84$^{th}$-50$^{th}$
of the metallicity estimate distribution.

The distributions for the D02 calibrator include the uncertainty in
the fit parameters: the oxygen abundance in this calibrator is
generated as \oxabinline~=~$9.12~\pm 0.05 + (0.73~\pm
0.10)\times\mathrm{[N~II]}/\ha$ \citep{denicolo02}. The parameters of
the fit are generated as the sum of the nominal parameters (9.12 and
0.73) and a Gaussian distributed random value centered on zero, and
within a standard deviation of 0.05 and 0.10, respectively, in the
above units.  Similarly, the distributions for the M13 calibrators
include the uncertainty in the fit parameters as stated in
\citet{marino13}: the oxygen abundance as a function of [N~II]/\ha~ is
parametrized as \oxabinline~=~$8.743~\pm 0.027 + (0.462~\pm
0.024)\times\mathrm{N~II}/\ha$, and as a function of
$\frac{\mathrm{[O~III]/H}\beta} {\mathrm{[N~II]/H}\alpha}$ as
\oxabinline~=~$8.533~\pm 0.012 + (0.214~\pm
0.012)\times\frac{\mathrm{[O~III]/H}\beta} {\mathrm{[N~II]/H}\alpha}$.

We note again that our code does not output the \emph{systematic}
uncertainty of each calibrator, which, for example, is $\sim$0.15
\nolinebreak dex for KD02. However, if all metallicity measurements
are in the \emph{same} calibrator and only \emph{relative} comparisons
are made, as recommended by a number of authors, then the systematic
error has no impact.



\subsubsection{Visual diagnostics to interpret the metallicity outputs}\label{sec:vizs}

\begin{deluxetable*}{lccccccccc} 
\tabletypesize{\tiny} \setlength{\tabcolsep}{0.0000001in}
\tablecolumns{10} \tablecaption{Example Data and their Uncertainties
  based on Data in \citet{modjaz11}} \tablehead{ 
  \colhead{site\tablenotemark{a}} & \colhead{[O~II]~$\lambda$3727} &
  \colhead{\hb} & \colhead{[O~III]~$\lambda$4959} &
  \colhead{[O~III]~$\lambda$5007} & \colhead{[O~I]~$\lambda$6300} &
  \colhead{\ha} & \colhead{[N~II]~$\lambda$6584} &
  \colhead{[S~II]~$\lambda$6717} & \colhead{[S~II]~$\lambda$6731} }
\startdata 08D & 1.842 (.053) & 0.958 (.032) & NaN & 0.302 (.029) &
0.127 (.021) & 4.746 (.026) & 1.642 (.022) & 0.941 (.021) & 0.543
(.019) \\ 06fo & 2.875 (.101) & 1.251 (.044) & 0.168 (.028) & 0.064
(.025) & NaN & 4.026 (.069) & 0.781 (.033) & 0.821 (.035) & 0.573
(.032) \enddata \tablenotetext{a}{These spectra were collected at the
  site of SN explosions: SN~2008D (line 1) and SN~2006fo (line 2)}
\label{tab:exampledata}
\end{deluxetable*}

\begin{figure}[ht!]
  \includegraphics[width=1.0\columnwidth]{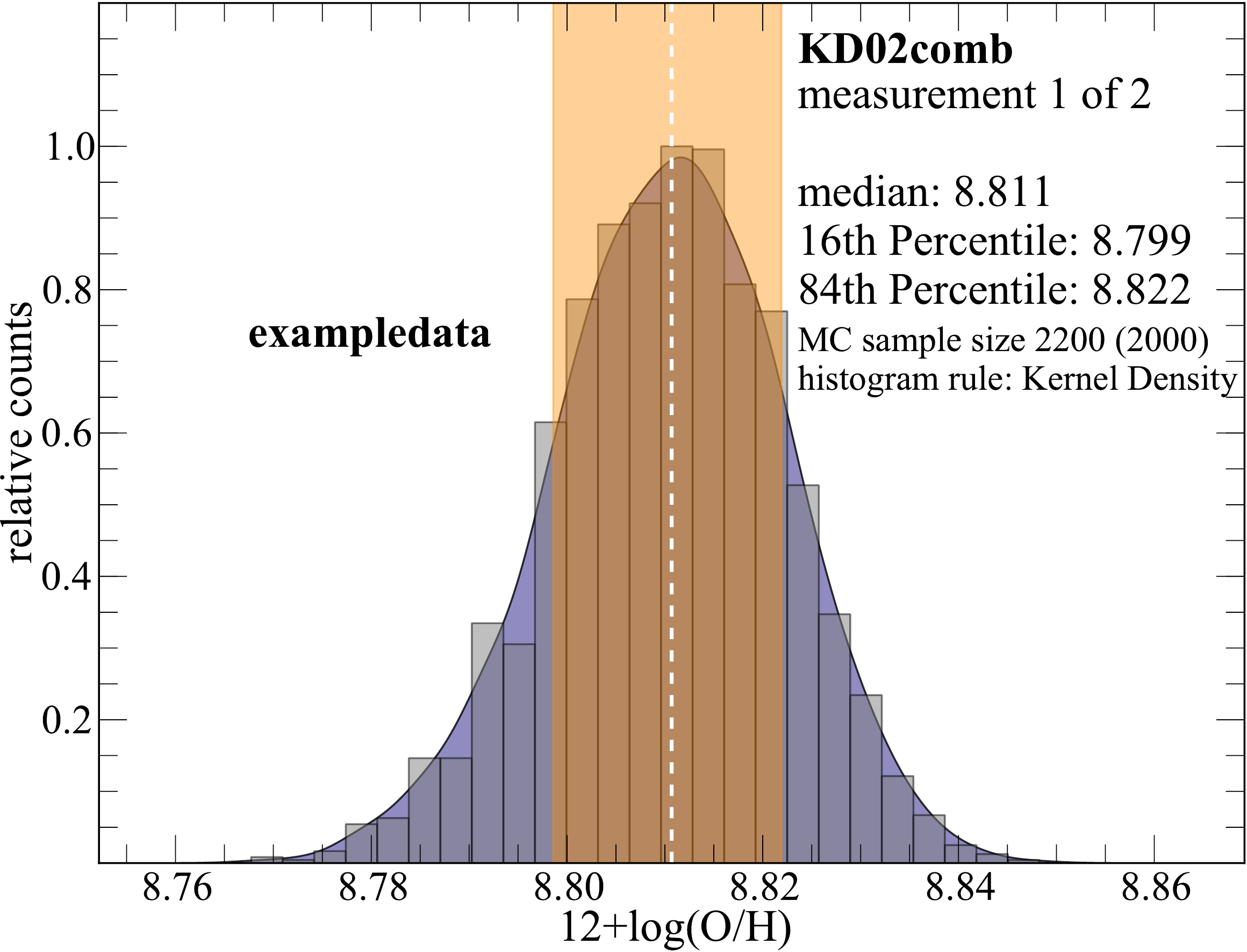}
   \caption{The Kernel Density for the distribution of values for the
     KD02comb calibrator for the first measurement of our example data
     (\autoref{tab:exampledata}). The Kernel Density is displayed as
     a blue shaded region, and is calculated via \emph{KD Tree} with a
     Gaussian kernel with bandwidth given by Silverman's rule, and
     normalized, as described in \autoref{sec:vizs}. The histogram of
     the distribution with bin size chosen according to Knuth's method
     is also plotted (gray bins) and the median and confidence
     intervals are shown as described in
     \autoref{fig:metallicity_distribution}.}\label{fig:KDE}
\end{figure}

\begin{figure}[ht!]
  \includegraphics[width=0.95\columnwidth]{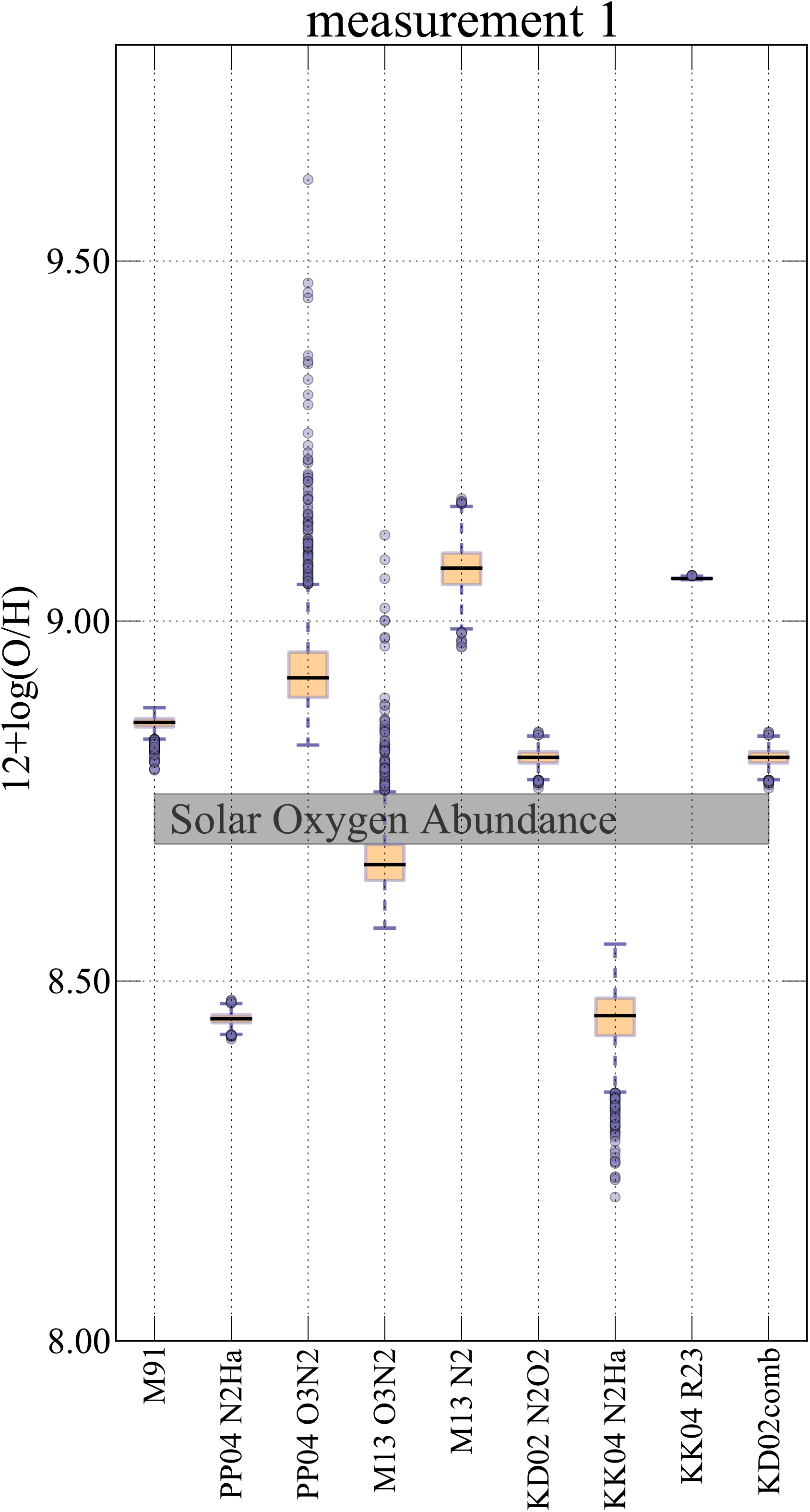}
   \caption{A box-and-whiskers plot showing the comparison of the
     results of nine metallicity calculations, corresponding to five
     calibrators as listed above, calculated from the same set of
     measured lines (\autoref{tab:exampledata}, host galaxy of
     SN~2008D). For each calibrator, the median of the resulting
     distribution is plotted as a horizontal line, the interquartile
     range (IQR) is represented as an orange box, and the bars, joined
     to each end of the box by a dashed line, represent the minimum
     and maximum of the distribution \emph{excluding outliers}, where
     outliers are defined as any point farther than $1.5\times$ IQR
     from the edges for the IQR. A range of values for Solar oxygen
     abundance that are commonly found in the literature is shown as
     a gray horizontal band.}
 \label{fig:boxplot}
\end{figure}

In order for the user to check the validity of a measurement, and to
better understand the distribution of metallicity values, we provide
two visualizations: for each set of input line fluxes, we generate a
histogram of the output distribution in all metallicity calibrators
calculated (\autoref{fig:metallicity_distribution},
\ref{fig:metallicity_bimodal}, and \ref{fig:KDE}), and for each set of
input line fluxes we generate a \emph{box-and-whiskers} plot
(hereafter \emph{boxplot}, for short) summarizing the result of all
calibrators calculated (\autoref{fig:boxplot}). All the plots generated
are created with Python
\verb=matplotlib=\footnote{\url{http://matplotlib.org/}}
\citep{hunter07}.

Choosing the binning size for a histogram is not a trivial task.
\citet{hogg08}, and \citet{astroMLText}, among others, describes
various data analysis recipes for selecting a histogram bin size. Too
many bins will result in an ``over-fit'' histogram (showing structure
that is in fact not in the distribution, thus perhaps decreasing the
confidence in the metallicity measurement), while too few bins may
miss features of the distribution (with the opposite effect).  We
provide several methods to generate the histograms, in order to
enable the best possible visualization, given the computational power,
for visual interpretation. However, we emphasize that our metallicity
and confidence interval calculation is robust to this choice, since it
calculates the percentiles of the unbinned distribution.

By default, we use \emph{Knuth's method} to choose the number of bins
$N_\mathrm{bins}$ for each histogram. Knuth's method optimizes a
Bayesian fitness function across fixed-width bins
\citep{knuth06}. Additionally, we enable a number of binning options
from which the user can choose, including: the square root of the
number of points, $N_\mathrm{bins}=\sqrt{N}$, \emph{Rice rule}
($N_\mathrm{bins}~=~2\sqrt[3]{N}$, e.g., \citealt{hastie09}),
\emph{Doane's formula} ($N_\mathrm{bins}~=~1 + \mathrm{log}_2{N} +
\mathrm{log}_2\left(1 + \mathrm{Kurt}\sqrt{(N / 6)}\right)$, where
Kurt is the third moment of the distribution,
\citealt{doane76}\footnote{\citet{doane76} attempted to address the
  issue of finding the proper number of bins for the histogram of a
  skewed distribution. Several versions of the so-called Doane's
  formula exist in the literature. The formula we implement is found
  in \citealt{bonate11}}), and the full Bayesian solution, known as
Bayesian Blocks, which optimizes a fitness function across an
arbitrary configuration of bins, such that the bins are of variable
size \citep{scargle13}. The implementation of the latter method
requires the
\verb=astroML=\footnote{\url{https://github.com/astroML/astroML}}
Python package to be installed on the user's system
(\citealt{astroml}). If the \verb=astroML= package is not found, the
code will default to Knuth's method.  As mentioned, Knuth's method
implies an optimization. In cases in which the convergence takes too
long, or if the number of bins after the minimization is
$N_\mathrm{bins}/\sqrt{N} > 5$ or $N_\mathrm{bins}/\sqrt{N} < 1/3$,
the code will revert to Rice rule.  Some methods may be
computationally prohibitive with a very large sample size, or very
little computational power, such as the Bayesian methods that rely on
optimization (Knuth's method and the Bayesian Block method) in which
case it may be convenient for a user to choose Doane's formula, Rice
rule, or even the square root of the number of samples.

Lastly, the user can generate and visualize the metallicity
distribution \emph{Kernel Density} if the
\verb=sklearn=\footnote{\url{http://scikit-learn.org/stable/}} package
is installed. Kernel Density Estimation (KDE) alleviates the problem
of choosing the bin size, at the cost of specifying a convolution
kernel. The Kernel Density of the distribution is here calculated via
\emph{KD Tree} with a gaussian kernel, as explained in the
\verb=sklearn= package
documentation.\footnote{\url{http://scikit-learn.org/stable/modules/density.html}}
The bandwidth of the kernel is chosen according to \emph{Silverman's
  rule} \citep{silverman86}.\footnote{By Silverman's rule the
  bandwidth for the KDE kernel is set to $w~=~1.06\sigma~N^{-0.2}$,
  where $\sigma$ is the sample standard deviation. Although the
  bandwidth chosen accordingly to Silverman's rule is only optimal in
  the case of a Gaussian basis and a Gaussian distribution, and our
  distributions are explicitly \emph{not} Gaussian, as explained
  earlier, this kernel parametrization generally provides good results
  for our metallicity distributions.} The results will then show both
a histogram, with $N_\mathrm{bins}$ chosen via Knuth's method, as well
as the distribution Kernel Density, as shown in
\autoref{fig:KDE}. The KDE is saved as a binary python object (an
\verb=sklearn= \verb=KernelDensity= object), so that it can be
recovered outside the code and used as a probability distribution for
further inference.

The histograms are always normalized so that the highest bin value is
1, and the Kernel Density is normalized to contain the same area as the
overplotted histogram.

At each run the code also generates a boxplot (\autoref{fig:boxplot})
that summarizes the result from each calibrator the user chooses to
calculate. For each calibrator, the median of the \oxabinline~
distribution is plotted as a black horizontal line. The height of the
corresponding box represents the $25^{th}$ percentile of the
\oxabinline~ distribution, known as the \emph{interquartile range}
(IQR).  The bars represent the maximum and minimum value of the
distribution, excluding outliers. The outliers are plotted as circles,
and are defined as all data points farther than
$1.5\times\mathrm{IQR}$ from the $25^{th}$ percentile (i.e. from
either end of the box).

The Solar oxygen abundance is indicated in this plot for comparison: a
gray box shows a range of estimated values for Solar oxygen
abundances, from \oxabinline=8.69 \citep{asplund09_rev} to
\oxabinline=8.76 \citep{chaffau11}.  The diagnostics requested by the
user have a slot in the plot (in the example in \autoref{fig:KDE}
the computed calibrators are the PP04, the M13, and the KD02
calibrators), as do diagnostics that are computed in support of other
requested diagnostics (M91 in this case, which is needed to calculate
KD02\_comb). However a slot exists in the plot whether the diagnostic
can be produced or not, i.e. if the set of input lines does not allow
a requested calibrator to be calculated an empty column will be
generated in this plot.

With this plot, the user can immediately check for consistency or
scatter in the metallicities derived by the requested calibrators
(e.g., there are well-known systematic offsets between different
methods, where the $T_e$-based, empirical methods usually give lower
values than the photoionization based ones), and coarsely assess the
shape of the distribution in each calibrator (e.g. strong asymmetry or
bimodality would show up as an asymmetric box, or a very asymmetric
distribution of outliers). Although, as we stated earlier, relative
comparison \emph{within} a metallicity calibrator are valid despite
the typical spread in metallicity derived by different strong line
methods, a particularly large spread may indicate difficulties in
assessing metallicity for this spectrum. The user is responsible for
understanding the strengths and caveats of the various diagnostics and
in which ranges and conditions they may be used. The visualization
tools we provide are meant to help the user understand the reliability
of the strong-line metallicity derived for a particular HII region.

\subsection{Visual diagnostic to assure completeness of the MC simulation}\label{sec:completeness}

\begin{figure*}[!ht]
\centerline{
  \includegraphics[width=0.65\columnwidth]{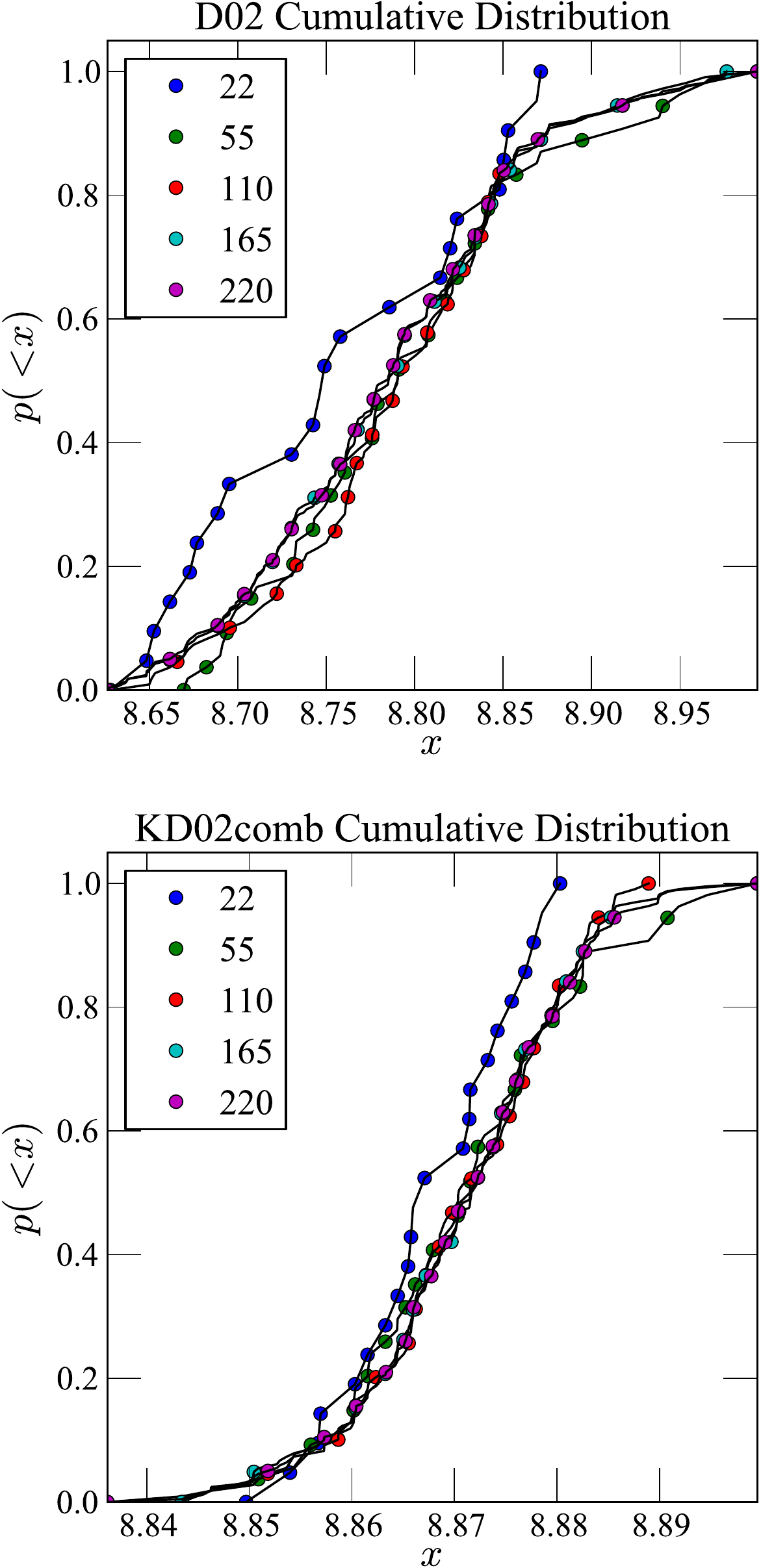}
  \includegraphics[width=0.65\columnwidth]{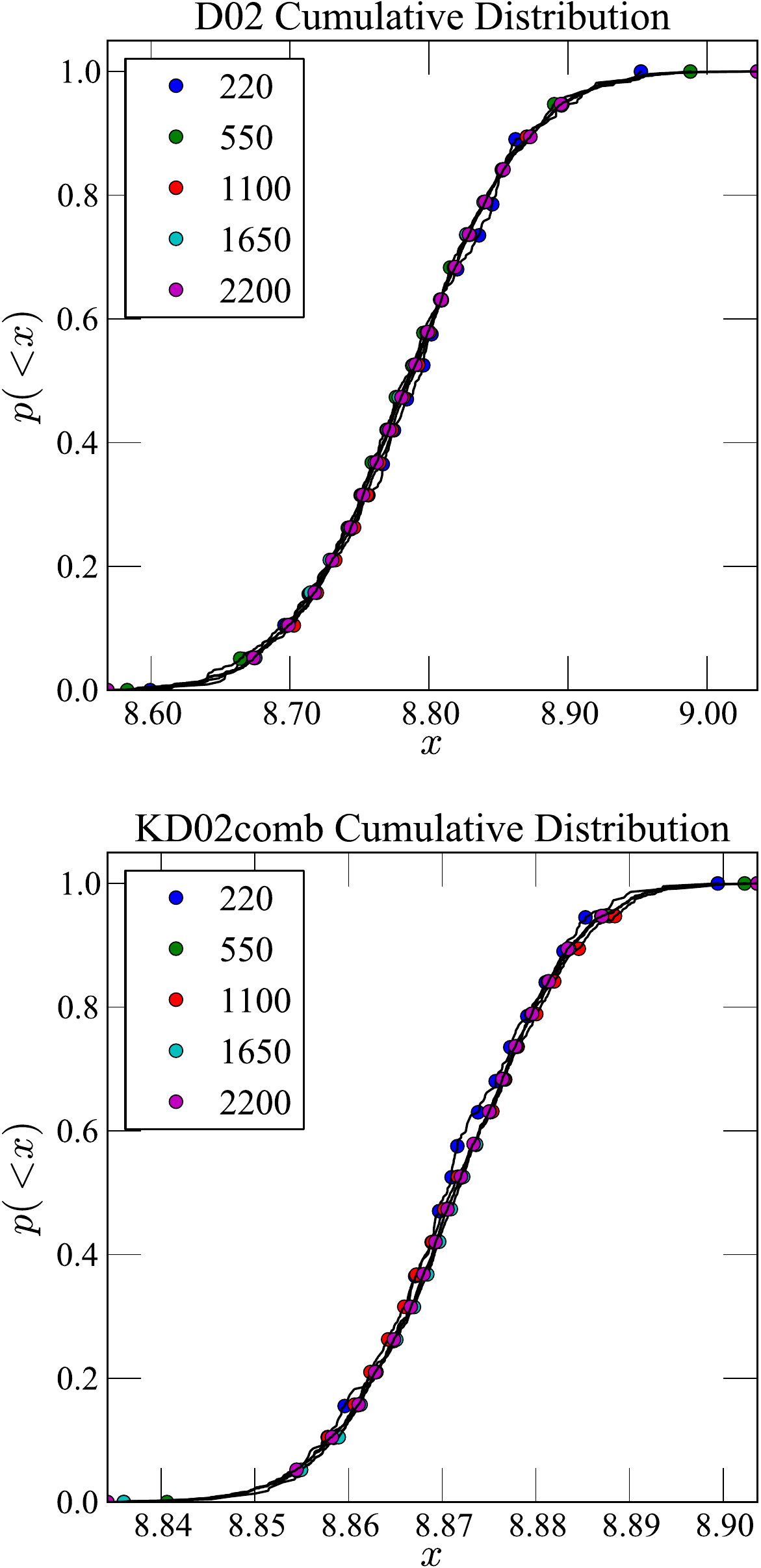}
  \includegraphics[width=0.65\columnwidth]{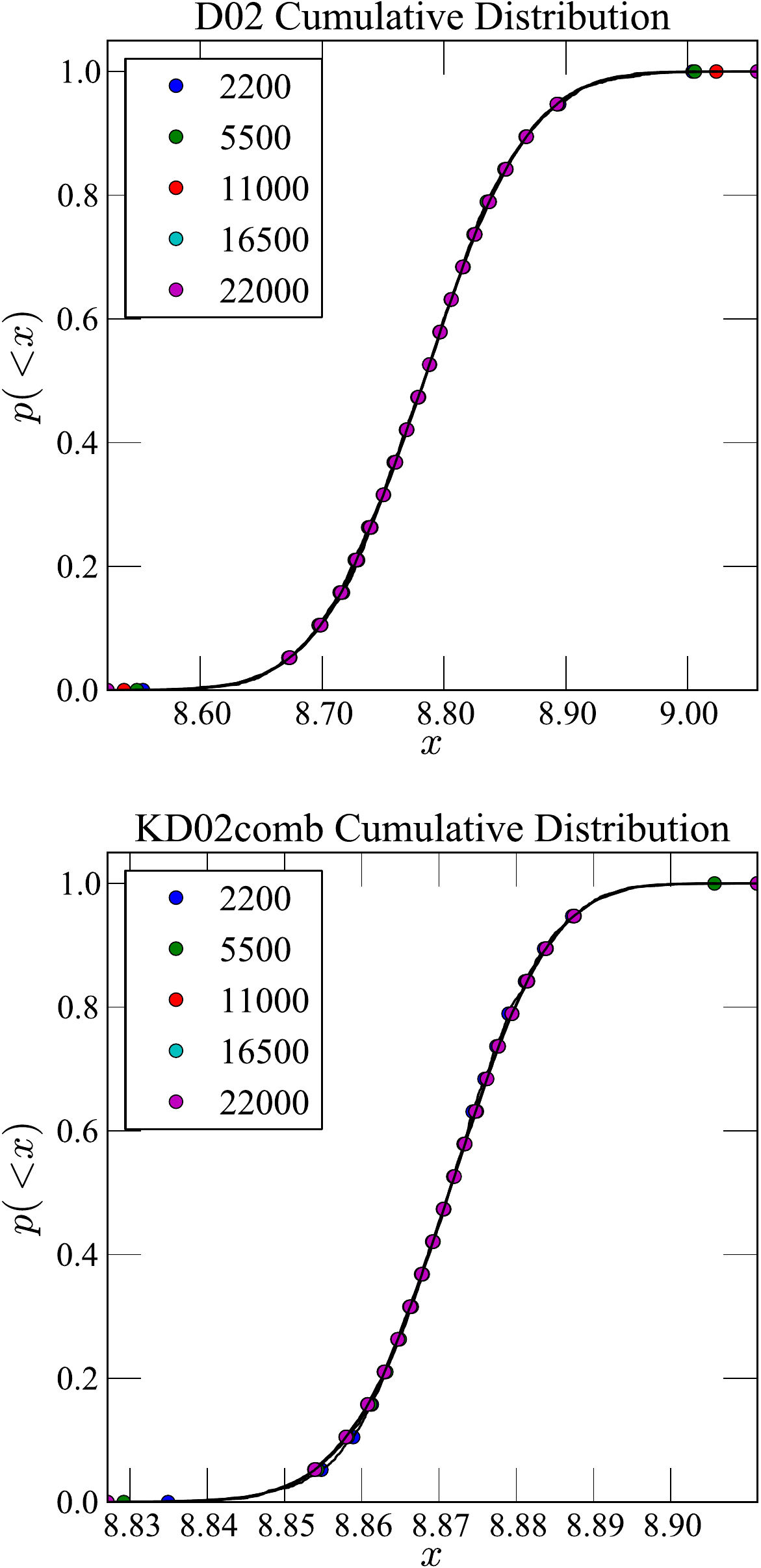}}
\caption{Cumulative plots of the distribution of metallicity values
  for the D02 \citep{denicolo02} and KD02 calibrator
  (\citealt{kewley02}, as updated by \citealt{kewley08}), chosen here
  just as examples, where $x$ indicates \oxabinline. The input data is
  ``example data 1'', the emission line values for the host galaxy of
  SN~2008D from \citet{modjaz11}. This plot provides a visual
  diagnostic of sample completeness. In each plot the cumulative
  distribution of metallicity values is shown for randomly chosen
  subsamples of 10\%, of 25\%, 50\%, and 75\% of the data, and for all
  data in the distribution. In the left top and bottom plots, the
  distributions are generated from an $N=200$ sample, in the center
  plots from an $N=2,000$, and in the right-most column from an
  $N=20,000$ samples. The increasing overlap of the distributions
  reflects increasing completeness. In the left plots, the
  distributions do not fully overlap, indicating that completeness is
  not achieved with the $N=200$ sample. On the other end, since all
  subsamples are indistinguishable in the rightmost top and bottom
  plots, from an $N=20,000$ sample, we conclude that completeness is
  already achieved at the smallest subsample in the plots on the right,
  10\% of the $N=20,000$ sample, for our example data. That is,
  $N=2,000$ is a sufficiently large sample for these data, and these
  diagnostics.}
 \label{fig:cd}
\end{figure*}

The user chooses $N$, the number of samples to be generated. The
sample size is automatically increased by 10\% at the beginning of the
run, in order to assure that even if during the calculations some of
the output metallicities were to result in non-valid values
(\emph{NaN}'s or infinities, for example, if divisions by very small
numbers are required) the actual sample size is at least as large as
the user intended.  The code rarely produces non-valid values, and if
the size of the valid output distribution were in the end smaller than
the \emph{requested} value $N$ (if the number of non-valid outputs is
larger than 0.1$N$), the user should worry about the reliability of
the set of input line fluxes used.

The reliability of the metallicity estimates depends crucially on the
sample size being sufficiently large to properly characterize the
distribution of metallicity values. It is, however, not trivial to
decide when $N$ is sufficiently large. \emph{As soon as $N$ is large
  enough, and the distribution is well characterized, increasing the
  synthetic sample size will not change its shape}.

Let us consider a cumulative distribution for a metallicity calibrator:
D02, for example, which has noise from the measurement errors, as well
as from the error in the fit parameters, or KD02, which uses a
non-linear combination of the input line flux values. We generate
three distributions with $N=200$, $N=2,000$, and $N=20,000$ for both
D02 and KD02comb.  For each of these distributions we randomly select
four subsamples of size 0.1$N$, 0.25$N$, 0.5$N$, and 0.75$N$, and we
overplot the cumulative distribution of each subsample to that of the
full sample. If all subsamples, including the smallest one (0.1$N$)
constituted a sufficiently large MC sample to properly characterize
the underlying distribution, all these cumulative distribution would
typically appear smooth and, most importantly, they would overlap.

In \autoref{fig:cd} we show these cumulative plots for a distribution
generated with $N=200$, $N=2,000$, and $N=20,000$ for D02 and
KD02comb. The cumulative distributions for the subsets of the $N=200$
samples simulation (left panel in \autoref{fig:cd}) are noisy and
rather different from each other, indicating that 200 data points are
not a sufficiently large dataset. Conversely, at $N=20,000$ (right
panel) the distributions are indistinguishable, indicating that a
sample of size $N=2,000$ (=0.1$N$) is already large enough to fully
describe the distribution. The $N=2,000$ cumulative distributions
rapidly converge as the subsample size increases, indicating that a
value of $N$ between 200 and 2,000 samples will characterize the
distribution appropriately for our example dataset.  In light of this
we choose, conservatively, $N=2,000$ to run our simulations for this
dataset.

The appropriate number $N$ of samples will depend on the calibrator
calculated and on the input data. The errors on the measurements and
the calculations performed to derive the metallicity from line ratios,
which for many calibrators are non-linear, will determine the shape of
the distribution, and thus the number of data points that are needed
to fully characterize it. Consult the repository
README\footnote{\url{https://github.com/nyusngroup/pyMCZ/blob/master/README.md}}
for instructions on how to generate these diagnostic plots.

\subsection{Performance and Benchmarks}
We ran a benchmark calculation on a MacBook Air with a dual-core Intel
Core i7 (1.7~GHz) and 8GB of 1600 MHz DDR3 memory. The dataset we used
for the calculation includes two sets of line measurements
(measurement 1, the host galaxy of SN~2008D, and measurement 2, the
host galaxy of SN~2006fo). The flux values and their associated errors
are shown in \autoref{tab:exampledata}. The code performs simple
algebraic operations on large data arrays. It is vectorized along the
sample-size dimension, so that the code loops over the smaller
dimension corresponding to the number of sets of line measurements in
input (two in our example data), while operations are generally
performed on the $N$-dimensional vectors storing the synthetic
measurements, and the $N$-dimensional variables derived from them
(certain calibrators, such as KD02 and KK04, derive the ionization
parameter in an iterative fashion, and require further loops).

With full graphic output (all histograms are plotted) and performing
all default metallicity calculations, except those of D13
(\emph{pyqz}), for our example datasets and a sample size $N=2,000$
the time required by the code is $\sim9$ seconds (wall clock time, and
less than half a second longer in total CPU time, as the machine we
tested on is dual-core), and less than 0.3 seconds of CPU time spent
in the kernel within the process. Including the calibrators of D13
\emph{pyqz} for the same datasets, the run time becomes $\sim19$
seconds of clock time (and actual CPU time).  The code time was tested
on sets of 1, 2, 5, 10, 25, 50, and 100 identical measurements (copies
of the emission line fluxes of the SN~2008D host galaxy in our example
dataset) and the clock time is found to scale roughly quadratically
with the number $m$ of measurement in input, but with a small
quadratic coefficient: $t~\sim~7m+0.07m^2$. This means effectively
that the 100 measurements sample will take 200 times longer than the 1
measurement sample (of course with dependence on which line values are
available in each measurement, enabling different metallicity
calibrators to be calculated). For the CPU time spent in the kernel we
find a roughly linear relation with the number of input measurements,
with a very shallow dependency: $t~\sim~ 0.05m$. This is the actual
computational time in calculations performed in the code: most clock
time is in fact spent in root finding, input-output, and plotting
activities.

\begin{figure}[ht!]
  \includegraphics[width=1.0\columnwidth]{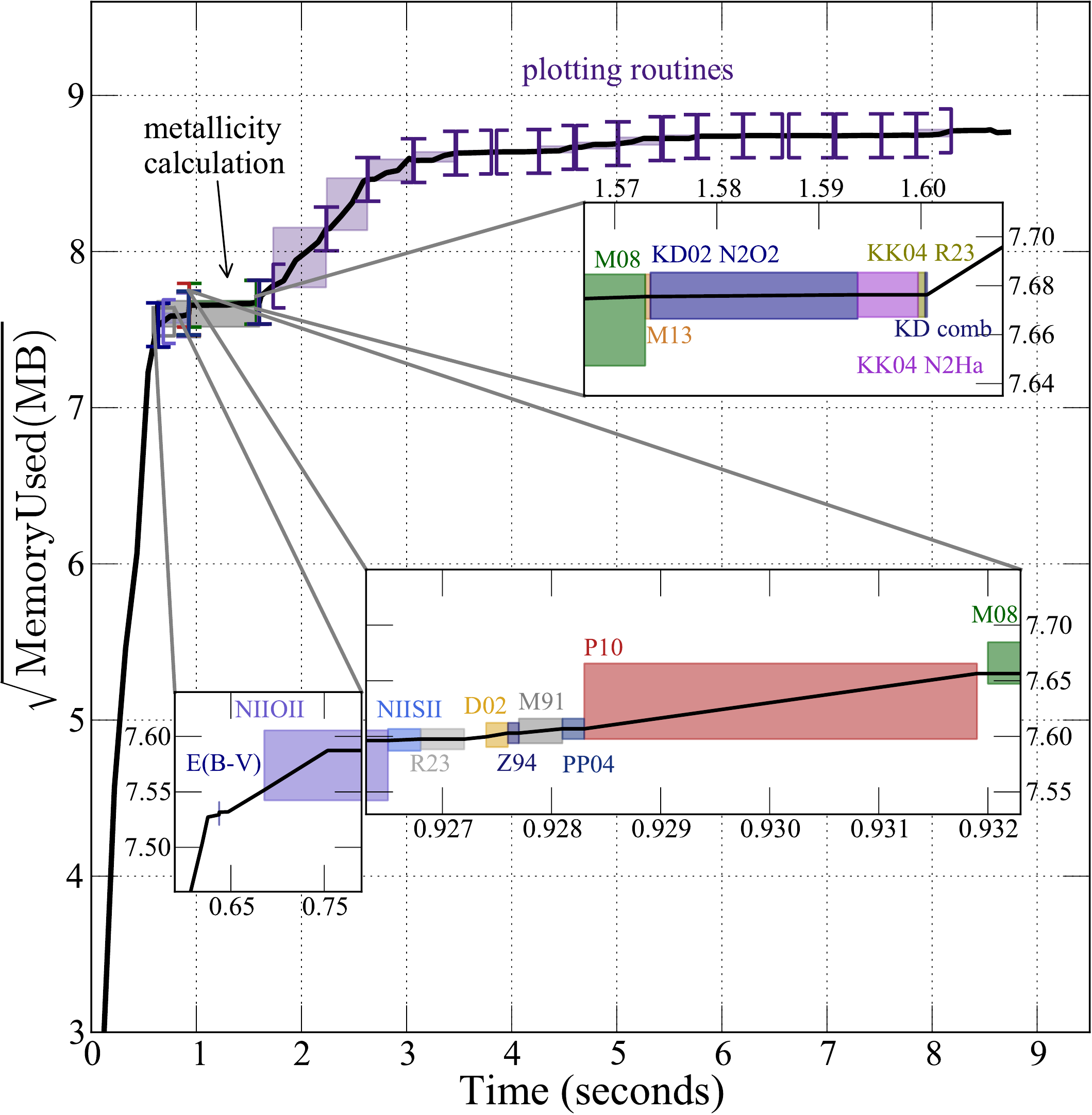}
   \caption{Memory usage: we plot the square root of the memory usage
     in Megabytes as a function of time for running our code (using
     $N$=2,000 and all default metallicity calibrators except the D13
     \emph{pyqz} ones, and PP04) on a single set of measured emission
     lines (\autoref{tab:exampledata}, host galaxy of SN~2008D). The
     square root is plotted, instead of the natural value, to enhance
     visibility.  Three inserts show the regions where most of the
     metallicity calibrators are calculated, zoomed in, since the run
     time of the code is dominated by plotting routines, including the
     calculation of the bin size with Knuth's rule.  Each function
     call is represented by an opening and closing bracket in the main
     plot, and by a shaded rectangle in the zoomed-in insets.  The
     calculation of $N2O2$, which requires 0.25 seconds, is split
     between two insets, as well as the calculation of the M08
     calibrators (three calibrators) which require 0.65
     seconds. Altogether, the calculation of metallicity in all
     default calibrators, except the D13 \emph{pyqz} calibrators,
     takes $\sim1$ seconds. For technical details about the D13
     \emph{pyqz} calibrators performance we refer the reader to the
     \emph{pyqz} package.}
 \label{fig:mem}
\end{figure}

The time spent on plotting functions, which includes the calculation
of the appropriate number of bins for each calibrator, is substantial:
1.67 seconds per distribution on average with 14 calls for this
dataset (one for each metallicity calibrator, \ebmv, and $R_{23}$).
We summarize the run time spent on each metallicity calibrator and
memory usage for the host galaxy of SN~2008D, the first measurement of
the sample dataset, in \autoref{fig:mem}. While the CPU usage is
modest, the memory usage can be large, depending on the size of the
sample.  The memory usage ramps up quickly, as soon as the $N$ line
samples are created, and remains fairly constant throughout the run.
\autoref{fig:mem} shows the memory used by each function in the code
as a function of time for a single set of input lines of our example
data (the host galaxy of SN~2008D). The calibrators that take longer
are those that require root finding (e.g. M08), or optimizations
(which are done iteratively, e.g. KD02\_N2O2).

The code is trivially parallelizable, as each measurement can be
computed on a different core, and multiprocessing is enabled via the
\verb=multiprocessing= Python module. When multiprocessing is enabled,
\verb=pyMCZ=it uses up to $n-1$ cores, where $n$ is the total number
of available cores to the user, or less, if a smaller maximum number
of cores is set by the user via the global variable MAXPROCESSES.

\section{Comparison to prior uncertainty computation and other works }\label{sec:comp_sec}

We compare our method with the results that appear in
\citealt{modjaz11} for 19 SN galaxies. We choose a SN host sample for
our comparison, since investigations of the environments of
extra-galactic transients, where observational errors dominate, are
the studies that can benefit most from our implementation of
metallicity calculations. We compare our results with the results that
appeared in \citealt{modjaz11} since for those results we have both
all flux values in input and their errors, and the metallicities and
uncertainties obtained in output. We note that this work includes only
three calibrators: M91, PP04\_O3N2, and KD02comb; however these
calibrators cover the case of calibrators that are fully linear, that
rely on decision trees, and mixed calibrators. We stress again that
the PP04 calibrators are superseded by the M13 ones, but they can be
used in this comparison nonetheless. Since our work is the
implementation of existing calculations, it is beyond the scope of
this paper to include a complete comparison of all calibrators: that
burden was on the original authors and we refer the reader to the
papers that present each method for that.

\begin{figure}[ht!]
  \includegraphics[width=1.0\columnwidth]{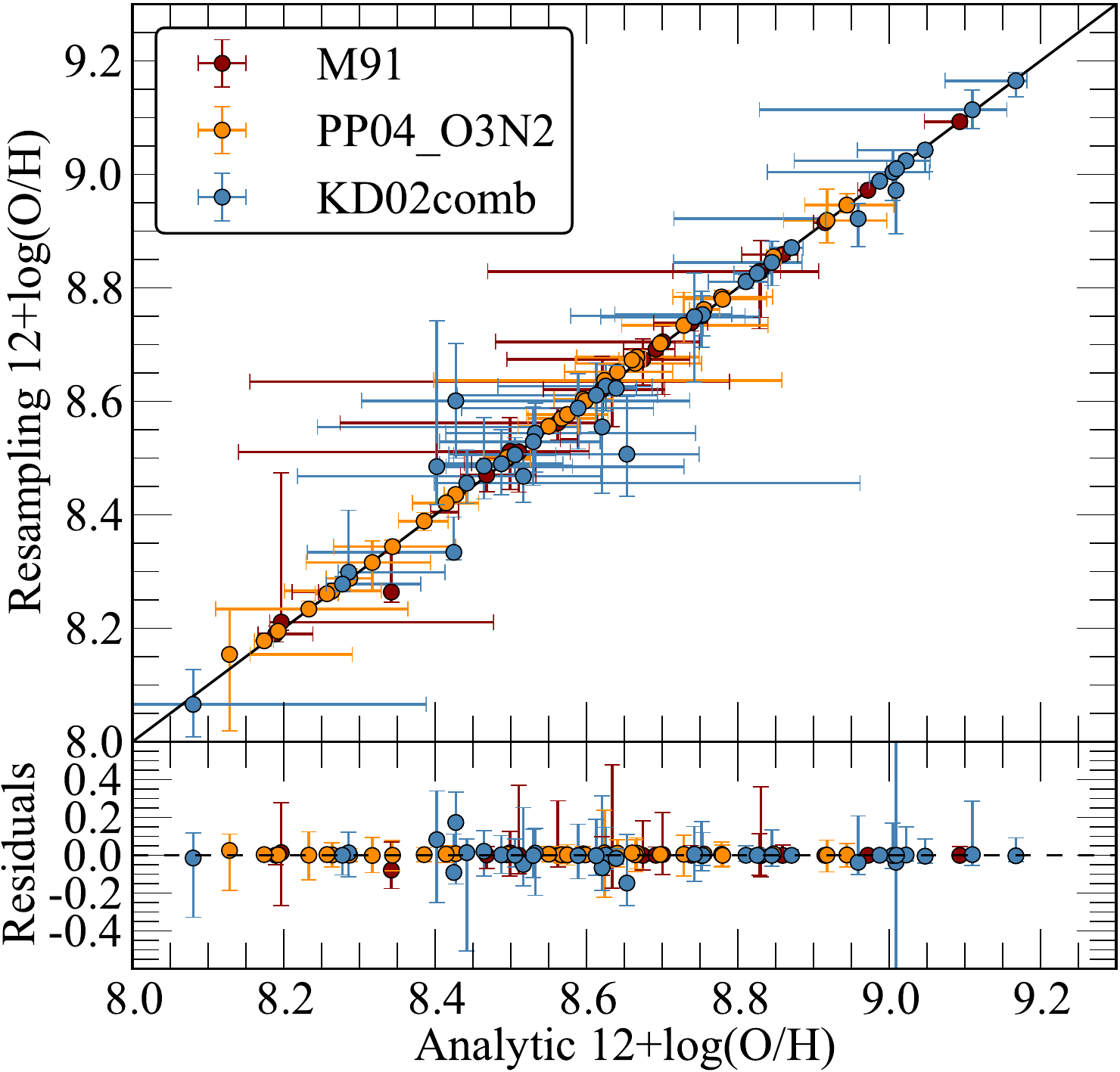}
   \caption{Comparison of metallicity estimation between the analytic
     method and our MC sampling method (top) and their residuals
     (bottom) for three different metallicity calibrators. Flux
     measurements come from 19 galaxies previously measured in
     \citet{modjaz11}.  Note that for KD02\_comb and M91, and in most
     cases for PP04\_O3N2 as well, the MC sampling error bars are
     smaller than those of the analytic propagation, which assumes
     worst-case scenarios. The metallicity value derived by our code,
     which is the 50th percentile of the metallicity distribution from
     MC sampling, is consistent with the analytically derived
     metallicity in all calibrators, for all measurements.}
 \label{fig:comp_anal_MC}
\end{figure}

A previous method for determining the uncertainty in the oxygen
abundance, used in \citealt{modjaz11} as well as, for example in
\citealt{modjaz08_Z,kewley10,rupke10} was an \emph{analytic} approach,
where the emission-line flux uncertainties are propagated into a
measurement ``envelope'': it found the extreme abundances obtained
from the input fluxes minimizing (maximizing) the indicators used in
the calculations by adding (subtracting) to the measured line flux
values their uncertainties, thus producing envelope values for the
metallicity. For comparison, we computed the metallicities and their
errors in both ways (analytically and using our current MC sampling
method) for three representative calibrators. We plot our results and
the residuals in \autoref{fig:comp_anal_MC}.  The errorbars in the
residuals are generated by adding asymmetric errors in quadrature:
$\mathrm{residual}_{\rm min}=\sqrt {x_{\rm max}^2 + y_{\rm min}^2}$
and $\mathrm{residual}_{\rm max}=\sqrt{x_{\rm min}^2 + y_{\rm
    max}^2}$.

This figure highlights a number of important points:
\begin{itemize}

\item
The metallicity reported as the $50^{th}$ percentile of the
metallicity parameter distribution from the MC sampling method is
consistent with the analytically derived metallicity. The mean of the
residuals is $\sim-2\times10^{-3}$, $\sim5\times10^{-3}$, and
$~\sim5\times10^{-2}$, for M91, PP04\_O3N2, and KD02\_comb
respectively, for 20, 32, and 35 valid measurement from the
\citet{modjaz11} data; the standard deviation of the residuals is
$\sim0.018,~ 0.005,~ 0.004$ respectively -- well within the respective
error bars - and thus, the published results still stand
(unsurprisingly, since our code, aside for the calculation of the
confidence interval, uses for these calibrators the same algorithms
developed for IDLKD02).
\item
The MC sampling method has smaller error bars than the analytic
method. This is easily understandable, since the analytic method
assumes the worst-case-scenarios, as it basically yields two
metallicity parameter draws (the ``minimum'' and ``maximum'') which
are in the tail of the full metallicity probability distribution. The
MC sampling method empirically characterizes the full parameter
estimation distribution.
\end{itemize}

The (minimal and statistically not significant) differences between
our and the IDLKD02 implementation of the KD02\_comb calibrator arise
from numerical differences in the solutions of the root finding
algorithms, and in the optimizations, which we carry out until
convergence, while IDLKD02 estimated three iterations would be
sufficient for minimization. The differences in the M91 and PP04\_O3O2
calibrators are only due to the shape of the distributions.

\subsection{Comparison with other works}

The field of SN host metallicity studies is rapidly developing, and
these studies are crucial avenues for constraining the progenitor
systems of different kinds of explosions. However, some the works in
this field do not compute errors in the derived metallicities, and
others do not show how they compute their statistical errors (e.g.,
\citealt{anderson10,leloudas11,sanders12,leloudas14}). 

In contrast, the general metallicity field has considered in detail
how to estimate the uncertainties in measured metallicities. However,
when the codes are designed for local Galaxy metallicity measurements,
they neglect the observational uncertainties, since they are
negligible compared to other sources of error. Other codes are not
open-source, and many of them are for specific calibrators. For
example: \citet{moustakas10} also use MC sampling to estimate the
metallicity uncertainties (in their case using $N$=500 trials and
assuming a Gaussian distribution for metallicity parameter
distribution) but only do this for two calibrators, KK04 and P05.

Some strong line methods fit a set of stellar population synthesis and
photoionization models directly to the spectra of HII regions by
changing the model parameters. For computing the metallicities of the
SDSS star forming galaxies, \citet{tremonti04} fit a combination of
stellar population synthesis models and photoionization models to the
observed strong emission lines [O~II], $\hb$, [O~III], $\ha$, [N~II]
and [S~II] and report the median of the metallicity likelihood
distribution as the metallicity estimate, with the width of the
distribution giving the 1-$\sigma$ (Gaussian) error.  \citet{blanc15}
use Bayesian inference to derive the joint and marginalized posterior
probability density functions for metallicity $Z$ and ionization
parameter $q$ given a set of observed line fluxes and an input
photoionization model. They provide a publicly available IDL
implementation of their method named $IZI$ (inferring metallicities --
$Z$ -- and ionization parameters).

\section{Conclusions}\label{sec:conclusions_sec}
We have presented the open-source Python code \verb=pyMCZ= that
determines oxygen abundance and its error distribution from strong
emission lines in a total of up to 15 diagnostics, 11 default and 4
additional upon request, where by diagnostic we intend a suite of
metallicity calibrators that are published in a single article, and
for a total of up to 32 different metallicity calibrators (since a
number of diagnostics produce different metallicity estimates based on
different input indicators). These estimates are based on the original
KD02IDL code in KD02, and updated in KE08, and expanded to include
more recently developed calibrators, and to generate confidence
intervals for each calibrator via MC sampling. We supply visualization
tools that, as part of the default output of \verb=pyMCZ=, enable the
user to assess the validity of each derived metallicity distribution
and understand when line flux measurements may lead to misleading
metallicity estimates, for example in proximity of the demarcation
between the upper and lower branch for $R_{23}$-based
methods. \verb=pyMCZ= outputs the full estimated metallicity
distribution (on demand), and its Kernel Density, and it offers
visualization tools to assess the spread of the oxygen abundance in
the different calibrators.

The validity of our metallicity measurements and of their confidence
regions hinges upon generating probability distributions that properly
sample the metallicity distribution, given the input parameters and
the specific metallicity calculation algorithm. Thus, we have
developed metrics that allow the user to ascertain that the sample
drawn in the simulation is sufficiently large. However, we note that
our code does not take into account systematic errors, but only
computes the statistical ones.

This code is open access and we welcome input and further development
from the community.  Since the code is open-source, the users can
include other metallicity calibrators and modify any parts of the
algorithms, or any assumption (e.g. that the line flux errors are
Gaussian distributed).

We hope that this open-access code will be helpful for the many
different fields where gas-phase metallicities are important, and
observational uncertainty are important in the error budget, including
in the emerging field of SN and GRB host galaxies.

\acknowledgements We thank John Moustakas for insightful
discussions. The Modjaz SNYU group at NYU is supported in parts by the
NSF CAREER award AST-1352405 and by NSF award
AST-1413260. F. B. Bianco is supported by a \emph{James Arthur
  Fellowship} at the NYU-Center for Cosmology and Particle Physics and
Y. Liu by a \emph{James Arthur Graduate Award}.  This code made use of
several Python Modules, including \verb=Matplotlib= .  Some plots were
produced with public code DOI:10.5281/zenodo.15419 available at
\url{https://github.com/fedhere/residuals_pylab}.  This research made
use of NASA Astrophysics Data System; the NASA/IPAC Extragalactic
Database (NED), which is operated by the Jet Propulsion Laboratory,
California Institute of Technology, under contract with the National
Aeronautics and Space Administration.

\bibliographystyle{apj}

\begin{thebibliography}{}
\expandafter\ifx\csname natexlab\endcsname\relax\def\natexlab#1{#1}\fi

\bibitem[{{Allen} {et~al.}(2008){Allen}, {Groves}, {Dopita}, {Sutherland}, \&
  {Kewley}}]{allen08}
{Allen}, M.~G., {Groves}, B.~A., {Dopita}, M.~A., {Sutherland}, R.~S., \&
  {Kewley}, L.~J. 2008, \apjs, 178, 20

\bibitem[{{Alloin} {et~al.}(1979){Alloin}, {Collin-Souffrin}, {Joly}, \&
  {Vigroux}}]{alloin79}
{Alloin}, D., {Collin-Souffrin}, S., {Joly}, M., \& {Vigroux}, L. 1979, \aap,
  78, 200

\bibitem[{{Anderson} {et~al.}(2010){Anderson}, {Covarrubias}, {James}, {Hamuy},
  \& {Habergham}}]{anderson10}
{Anderson}, J.~P., {Covarrubias}, R.~A., {James}, P.~A., {Hamuy}, M., \&
  {Habergham}, S.~M. 2010, \mnras, 407, 2660

\bibitem[{{Andrae}(2010)}]{andrae10}
{Andrae}, R. 2010, ArXiv e-prints, arXiv:1009.2755

\bibitem[{{Asplund} {et~al.}(2009){Asplund}, {Grevesse}, {Sauval}, \&
  {Scott}}]{asplund09_rev}
{Asplund}, M., {Grevesse}, N., {Sauval}, A.~J., \& {Scott}, P. 2009, \araa, 47,
  481

\bibitem[{{Baldwin} {et~al.}(1981){Baldwin}, {Phillips}, \&
  {Terlevich}}]{baldwin81}
{Baldwin}, J.~A., {Phillips}, M.~M., \& {Terlevich}, R. 1981, \pasp, 93, 5

\bibitem[{{Berg} {et~al.}(2015){Berg}, {Croxall}, {Skillman}, {Pogge},
  {Moustakas}, \& {Groh-Johnson}}]{berg15}
{Berg}, D.~A., {Croxall}, K.~V., {Skillman}, E.~D., {et~al.} 2015, \apj, 808, 42.

\bibitem[{{Blanc} {et~al.}(2015){Blanc}, {Kewley}, {Vogt}, \&
  {Dopita}}]{blanc15}
{Blanc}, G.~A., {Kewley}, L., {Vogt}, F.~P.~A., \& {Dopita}, M.~A. 2015, \apj,
  798, 99

\bibitem[{Bonate(2011)}]{bonate11}
Bonate, P. 2011, Pharmacokinetic-Pharmacodynamic Modeling and Simulation,
  SpringerLink : B{\"u}cher (Springer)

\bibitem[{{Bruzual A.} \& {Charlot}(1993)}]{bruzual93}
{Bruzual A.}, G., \& {Charlot}, S. 1993, \apj, 405, 538

\bibitem[{{Bundy} {et~al.}(2015){Bundy}, {Bershady}, {Law}, {Yan}, {Drory},
  {MacDonald}, {Wake}, {Cherinka}, {S{\'a}nchez-Gallego}, {Weijmans}, {Thomas},
  {Tremonti}, {Masters}, {Coccato}, {Diamond-Stanic}, {Arag{\'o}n-Salamanca},
  {Avila-Reese}, {Badenes}, {Falc{\'o}n-Barroso}, {Belfiore}, {Bizyaev},
  {Blanc}, {Bland-Hawthorn}, {Blanton}, {Brownstein}, {Byler}, {Cappellari},
  {Conroy}, {Dutton}, {Emsellem}, {Etherington}, {Frinchaboy}, {Fu}, {Gunn},
  {Harding}, {Johnston}, {Kauffmann}, {Kinemuchi}, {Klaene}, {Knapen},
  {Leauthaud}, {Li}, {Lin}, {Maiolino}, {Malanushenko}, {Malanushenko}, {Mao},
  {Maraston}, {McDermid}, {Merrifield}, {Nichol}, {Oravetz}, {Pan}, {Parejko},
  {Sanchez}, {Schlegel}, {Simmons}, {Steele}, {Steinmetz}, {Thanjavur},
  {Thompson}, {Tinker}, {van den Bosch}, {Westfall}, {Wilkinson}, {Wright},
  {Xiao}, \& {Zhang}}]{Bundy15}
{Bundy}, K., {Bershady}, M.~A., {Law}, D.~R., {et~al.} 2015, \apj, 798, 7

\bibitem[{{Caffau} {et~al.}(2011){Caffau}, {Ludwig}, {Steffen}, {Freytag}, \&
  {Bonifacio}}]{chaffau11}
{Caffau}, E., {Ludwig}, H.-G., {Steffen}, M., {Freytag}, B., \& {Bonifacio}, P.
  2011, \solphys, 268, 255

\bibitem[{{Cardelli} {et~al.}(1989){Cardelli}, {Clayton}, \&
  {Mathis}}]{cardelli89}
{Cardelli}, J.~A., {Clayton}, G.~C., \& {Mathis}, J.~S. 1989, \apj, 345, 245

\bibitem[{{Charlot} \& {Longhetti}(2001)}]{charlot01}
{Charlot}, S., \& {Longhetti}, M. 2001, \mnras, 323, 887

\bibitem[{{Cid Fernandes} {et~al.}(2010){Cid Fernandes}, {Stasi{\'n}ska},
  {Schlickmann}, {Mateus}, {Vale Asari}, {Schoenell}, \&
  {Sodr{\'e}}}]{cidfernandes10}
{Cid Fernandes}, R., {Stasi{\'n}ska}, G., {Schlickmann}, M.~S., {et~al.} 2010,
  \mnras, 403, 1036

\bibitem[{{Denicol{\'o}} {et~al.}(2002){Denicol{\'o}}, {Terlevich}, \&
  {Terlevich}}]{denicolo02}
{Denicol{\'o}}, G., {Terlevich}, R., \& {Terlevich}, E. 2002, \mnras, 330, 69

\bibitem[{{D{\'{\i}}az} \& {P{\'e}rez-Montero}(2000)}]{diaz00}
{D{\'{\i}}az}, A.~I., \& {P{\'e}rez-Montero}, E. 2000, \mnras, 312, 130

\bibitem[{Doane(1976)}]{doane76}
Doane, D.~P. 1976, The American Statistician, 30, 181

\bibitem[{{Dopita} \& {Evans}(1986)}]{dopita86}
{Dopita}, M.~A., \& {Evans}, I.~N. 1986, \apj, 307, 431

\bibitem[{{Dopita} {et~al.}(2013){Dopita}, {Sutherland}, {Nicholls}, {Kewley},
  \& {Vogt}}]{dopita13}
{Dopita}, M.~A., {Sutherland}, R.~S., {Nicholls}, D.~C., {Kewley}, L.~J., \&
  {Vogt}, F.~P.~A. 2013, \apjs, 208, 10

\bibitem[{{Efron}(1979)}]{efron79}
{Efron}, R. 1979, Ann. Stat., 7, 1

\bibitem[{{Evans} \& {Dopita}(1985)}]{evans85}
{Evans}, I.~N., \& {Dopita}, M.~A. 1985, \apjs, 58, 125

\bibitem[{{Ferland} {et~al.}(1998){Ferland}, {Korista}, {Verner}, {Ferguson},
  {Kingdon}, \& {Verner}}]{ferland98}
{Ferland}, G.~J., {Korista}, K.~T., {Verner}, D.~A., {et~al.} 1998, \pasp, 110,
  761

\bibitem[{{Grevesse} {et~al.}(2010){Grevesse}, {Asplund}, {Sauval}, \&
  {Scott}}]{grevesse10}
{Grevesse}, N., {Asplund}, M., {Sauval}, A.~J., \& {Scott}, P. 2010, \apss,
  328, 179

\bibitem[{{Hastie} {et~al.}(2009){Hastie}, {Tibshirani}, \&
  {Friedman}}]{hastie09}
{Hastie}, T., {Tibshirani}, R., \& {Friedman}, J. 2009, {The Elements of
  Statistical Learning: Data Mining, Inference, and Prediction} (Springer
  Science+Business Media, New York)

\bibitem[{{Hogg}(2008)}]{hogg08}
{Hogg}, D.~W. 2008, ArXiv e-prints, arXiv:0807.4820

\bibitem[{Hunter(2007)}]{hunter07}
Hunter, J.~D. 2007, Computing In Science \& Engineering, 9, 90

\bibitem[{{Ivezi{\'c}} {et~al.}(2014){Ivezi{\'c}}, {Connolly}, {Vanderplas}, \&
  {Gray}}]{astroMLText}
{Ivezi{\'c}}, {\v Z}., {Connolly}, A., {Vanderplas}, J., \& {Gray}, A. 2014,
  Statistics, Data Mining and Machine Learning in Astronomy (Princeton
  University Press)

\bibitem[{{Johnson} \& {Li}(2012)}]{johnson12}
{Johnson}, J.~L., \& {Li}, H. 2012, \apj, 751, 81

\bibitem[{{Kauffmann} {et~al.}(2003){Kauffmann}, {Heckman}, {Tremonti},
  {Brinchmann}, {Charlot}, {White}, {Ridgway}, {Brinkmann}, {Fukugita}, {Hall},
  {Ivezi{\'c}}, {Richards}, \& {Schneider}}]{kauffmann03}
{Kauffmann}, G., {Heckman}, T.~M., {Tremonti}, C., {et~al.} 2003, \mnras, 346,
  1055

\bibitem[{{Kelly} \& {Kirshner}(2012)}]{kelly12}
{Kelly}, P.~L., \& {Kirshner}, R.~P. 2012, \apj, 759, 107

\bibitem[{{Kewley} \& {Dopita}(2002)}]{kewley02}
{Kewley}, L.~J., \& {Dopita}, M.~A. 2002, \apjs, 142, 35

\bibitem[{{Kewley} \& {Ellison}(2008)}]{kewley08}
{Kewley}, L.~J., \& {Ellison}, S.~L. 2008, \apj, 681, 1183

\bibitem[{{Kewley} {et~al.}(2006){Kewley}, {Groves}, {Kauffmann}, \&
  {Heckman}}]{kewley06_sdss}
{Kewley}, L.~J., {Groves}, B., {Kauffmann}, G., \& {Heckman}, T. 2006, \mnras,
  372, 961

\bibitem[{{Kewley} {et~al.}(2010){Kewley}, {Rupke}, {Zahid}, {Geller}, \&
  {Barton}}]{kewley10}
{Kewley}, L.~J., {Rupke}, D., {Zahid}, H.~J., {Geller}, M.~J., \& {Barton},
  E.~J. 2010, \apjl, 721, L48

\bibitem[{{Knuth}(2006)}]{knuth06}
{Knuth}, K.~H. 2006, ArXiv Physics e-prints, physics/0605197

\bibitem[{{Kobulnicky} \& {Kewley}(2004)}]{kobulnicky04}
{Kobulnicky}, H.~A., \& {Kewley}, L.~J. 2004, \apj, 617, 240

\bibitem[{{Leloudas} {et~al.}(2011){Leloudas}, {Gallazzi}, {Sollerman},
  {Stritzinger}, {Fynbo}, {Hjorth}, {Malesani}, {Micha{\l}owski},
  {Milvang-Jensen}, \& {Smith}}]{leloudas11}
{Leloudas}, G., {Gallazzi}, A., {Sollerman}, J., {et~al.} 2011, \aap, 530, A95

\bibitem[{{Leloudas} {et~al.}(2014){Leloudas}, {Schulze}, {Kruehler},
  {Gorosabel}, {Christensen}, {Mehner}, {de Ugarte Postigo}, {Amorin},
  {Thoene}, {Anderson}, {Bauer}, {Gallazzi}, {Helminiak}, {Hjorth}, {Ibar},
  {Malesani}, {Morell}, {Vinko}, \& {Wheeler}}]{leloudas14}
{Leloudas}, G., {Schulze}, S., {Kruehler}, T., {et~al.} 2015, \mnras, 449, 917L. 

\bibitem[{{Levesque} {et~al.}(2010){Levesque}, {Berger}, {Kewley}, \&
  {Bagley}}]{levesque10_grbhosts}
{Levesque}, E.~M., {Berger}, E., {Kewley}, L.~J., \& {Bagley}, M.~M. 2010, \aj,
  139, 694

\bibitem[{{L{\'o}pez-S{\'a}nchez} {et~al.}(2012){L{\'o}pez-S{\'a}nchez},
  {Dopita}, {Kewley}, {Zahid}, {Nicholls}, \&
  {Scharw{\"a}chter}}]{lopezsanchez12}
{L{\'o}pez-S{\'a}nchez}, {\'A}.~R., {Dopita}, M.~A., {Kewley}, L.~J., {et~al.}
  2012, \mnras, 426, 2630

\bibitem[{{Lunnan} {et~al.}(2014){Lunnan}, {Chornock}, {Berger}, {Laskar},
  {Fong}, {Rest}, {Sanders}, {Challis}, {Drout}, {Foley}, {Huber}, {Kirshner},
  {Leibler}, {Marion}, {McCrum}, {Milisavljevic}, {Narayan}, {Scolnic},
  {Smartt}, {Smith}, {Soderberg}, {Tonry}, {Burgett}, {Chambers}, {Flewelling},
  {Hodapp}, {Kaiser}, {Magnier}, {Price}, \& {Wainscoat}}]{lunnan14}
{Lunnan}, R., {Chornock}, R., {Berger}, E., {et~al.} 2014, \apj, 787, 138

\bibitem[{{Maiolino} {et~al.}(2008){Maiolino}, {Nagao}, {Grazian}, {Cocchia},
  {Marconi}, {Mannucci}, {Cimatti}, {Pipino}, {Ballero}, {Calura}, {Chiappini},
  {Fontana}, {Granato}, {Matteucci}, {Pastorini}, {Pentericci}, {Risaliti},
  {Salvati}, \& {Silva}}]{maiolino08}
{Maiolino}, R., {Nagao}, T., {Grazian}, A., {et~al.} 2008, \aap, 488, 463

\bibitem[{{Marino} {et~al.}(2013){Marino}, {Rosales-Ortega}, {S{\'a}nchez},
  {Gil de Paz}, {V{\'{\i}}lchez}, {Miralles-Caballero}, {Kehrig},
  {P{\'e}rez-Montero}, {Stanishev}, {Iglesias-P{\'a}ramo}, {D{\'{\i}}az},
  {Castillo-Morales}, {Kennicutt}, {L{\'o}pez-S{\'a}nchez}, {Galbany},
  {Garc{\'{\i}}a-Benito}, {Mast}, {Mendez-Abreu}, {Monreal-Ibero}, {Husemann},
  {Walcher}, {Garc{\'{\i}}a-Lorenzo}, {Masegosa}, {Del Olmo Orozco},
  {Mour{\~a}o}, {Ziegler}, {Moll{\'a}}, {Papaderos},
  {S{\'a}nchez-Bl{\'a}zquez}, {Gonz{\'a}lez Delgado}, {Falc{\'o}n-Barroso},
  {Roth}, {van de Ven}, \& {Califa Team}}]{marino13}
{Marino}, R.~A., {Rosales-Ortega}, F.~F., {S{\'a}nchez}, S.~F., {et~al.} 2013,
  \aap, 559, A114

\bibitem[{{McGaugh}(1991)}]{mcgaugh91}
{McGaugh}, S.~S. 1991, \apj, 380, 140

\bibitem[{Mendoza \& Bautista(2014)}]{Mendoza14}
Mendoza, C., \& Bautista, M.~A. 2014, The Astrophysical Journal, 785, 91

\bibitem[{{Modjaz}(2012)}]{modjaz12_proc}
{Modjaz}, M. 2012, in IAU Symposium, Vol. 279, IAU Symposium, 207--211

\bibitem[{{Modjaz} {et~al.}(2011){Modjaz}, {Kewley}, {Bloom}, {Filippenko},
  {Perley}, \& {Silverman}}]{modjaz11}
{Modjaz}, M., {Kewley}, L., {Bloom}, J.~S., {et~al.} 2011, \apjl, 731, L4

\bibitem[{{Modjaz} {et~al.}(2008){Modjaz}, {Kewley}, {Kirshner}, {Stanek},
  {Challis}, {Garnavich}, {Greene}, {Kelly}, \& {Prieto}}]{modjaz08_Z}
{Modjaz}, M., {Kewley}, L., {Kirshner}, R.~P., {et~al.} 2008, \aj, 135, 1136

\bibitem[{{Moll{\'a}} {et~al.}(2009){Moll{\'a}}, {Garc{\'{\i}}a-Vargas}, \&
  {Bressan}}]{molla09}
{Moll{\'a}}, M., {Garc{\'{\i}}a-Vargas}, M.~L., \& {Bressan}, A. 2009, \mnras,
  398, 451

\bibitem[{{Moustakas} {et~al.}(2010){Moustakas}, {Kennicutt}, {Tremonti},
  {Dale}, {Smith}, \& {Calzetti}}]{moustakas10}
{Moustakas}, J., {Kennicutt}, Jr., R.~C., {Tremonti}, C.~A., {et~al.} 2010,
  \apjs, 190, 233

\bibitem[{{Nicholls} {et~al.}(2012){Nicholls}, {Dopita}, \&
  {Sutherland}}]{nicholls12}
{Nicholls}, D.~C., {Dopita}, M.~A., \& {Sutherland}, R.~S. 2012, \apj, 752, 148

\bibitem[{{Osterbrock}(1989)}]{osterbrock89}
{Osterbrock}, D.~E. 1989, {Astrophysics of Gaseous Nebulae and Active Galaxies}
  (Mill Vallery: University Science Books)

\bibitem[{{Owocki} \& {Scudder}(1983)}]{owoki83}
{Owocki}, S.~P., \& {Scudder}, J.~D. 1983, \apj, 270, 758

\bibitem[{{Pagel} {et~al.}(1979){Pagel}, {Edmunds}, {Blackwell}, {Chun}, \&
  {Smith}}]{pagel79}
{Pagel}, B.~E.~J., {Edmunds}, M.~G., {Blackwell}, D.~E., {Chun}, M.~S., \&
  {Smith}, G. 1979, \mnras, 189, 95

\bibitem[{{Pan} {et~al.}(2014){Pan}, {Sullivan}, {Maguire}, {Hook}, {Nugent},
  {Howell}, {Arcavi}, {Botyanszki}, {Cenko}, {DeRose}, {Fakhouri}, {Gal-Yam},
  {Hsiao}, {Kulkarni}, {Laher}, {Lidman}, {Nordin}, {Walker}, \& {Xu}}]{pan14}
{Pan}, Y.-C., {Sullivan}, M., {Maguire}, K., {et~al.} 2014, \mnras, 438, 1391

\bibitem[{{Peimbert}(1967)}]{peimbert67}
{Peimbert}, M. 1967, \apj, 150, 825

\bibitem[{{P{\'e}rez-Montero}(2014)}]{perezmontero14}
{P{\'e}rez-Montero}, E. 2014, \mnras, 441, 2663

\bibitem[{{Pettini} \& {Pagel}(2004)}]{pettini04}
{Pettini}, M., \& {Pagel}, B.~E.~J. 2004, \mnras, 348, L59

\bibitem[{{Pilyugin}(2001)}]{pilyugin01}
{Pilyugin}, L.~S. 2001, \aap, 369, 594

\bibitem[{{Pilyugin} \& {Thuan}(2005)}]{pilyugin05}
{Pilyugin}, L.~S., \& {Thuan}, T.~X. 2005, \apj, 631, 231

\bibitem[{{Pilyugin} {et~al.}(2012){Pilyugin}, {V{\'{\i}}lchez}, {Mattsson}, \&
  {Thuan}}]{pilyugin12}
{Pilyugin}, L.~S., {V{\'{\i}}lchez}, J.~M., {Mattsson}, L., \& {Thuan}, T.~X.
  2012, \mnras, 421, 1624

\bibitem[{{Pilyugin} {et~al.}(2010){Pilyugin}, {V{\'{\i}}lchez}, \&
  {Thuan}}]{pilyugin10}
{Pilyugin}, L.~S., {V{\'{\i}}lchez}, J.~M., \& {Thuan}, T.~X. 2010, \apj, 720,
  1738

\bibitem[{{Rupke} {et~al.}(2010){Rupke}, {Kewley}, \& {Chien}}]{rupke10}
{Rupke}, D.~S.~N., {Kewley}, L.~J., \& {Chien}, L.-H. 2010, \apj, 723, 1255

\bibitem[{{S{\'a}nchez} {et~al.}(2012){S{\'a}nchez}, {Kennicutt}, {Gil de Paz},
  {van de Ven}, {V{\'{\i}}lchez}, {Wisotzki}, {Walcher}, {Mast}, {Aguerri},
  {Albiol-P{\'e}rez}, {Alonso-Herrero}, {Alves}, {Bakos}, {Bart{\'a}kov{\'a}},
  {Bland-Hawthorn}, {Boselli}, {Bomans}, {Castillo-Morales}, {Cortijo-Ferrero},
  {de Lorenzo-C{\'a}ceres}, {Del Olmo}, {Dettmar}, {D{\'{\i}}az}, {Ellis},
  {Falc{\'o}n-Barroso}, {Flores}, {Gallazzi}, {Garc{\'{\i}}a-Lorenzo},
  {Gonz{\'a}lez Delgado}, {Gruel}, {Haines}, {Hao}, {Husemann},
  {Igl{\'e}sias-P{\'a}ramo}, {Jahnke}, {Johnson}, {Jungwiert}, {Kalinova},
  {Kehrig}, {Kupko}, {L{\'o}pez-S{\'a}nchez}, {Lyubenova}, {Marino},
  {M{\'a}rmol-Queralt{\'o}}, {M{\'a}rquez}, {Masegosa}, {Meidt},
  {Mendez-Abreu}, {Monreal-Ibero}, {Montijo}, {Mour{\~a}o}, {Palacios-Navarro},
  {Papaderos}, {Pasquali}, {Peletier}, {P{\'e}rez}, {P{\'e}rez}, {Quirrenbach},
  {Rela{\~n}o}, {Rosales-Ortega}, {Roth}, {Ruiz-Lara},
  {S{\'a}nchez-Bl{\'a}zquez}, {Sengupta}, {Singh}, {Stanishev}, {Trager},
  {Vazdekis}, {Viironen}, {Wild}, {Zibetti}, \& {Ziegler}}]{sanchez12}
{S{\'a}nchez}, S.~F., {Kennicutt}, R.~C., {Gil de Paz}, A., {et~al.} 2012,
  \aap, 538, A8

\bibitem[{{S{\'a}nchez} {et~al.}(2015){S{\'a}nchez}, {P{\'e}rez},
  {Rosales-Ortega}, {Miralles-Caballero}, {L{\'o}pez-S{\'a}nchez},
  {Iglesias-P{\'a}ramo}, {Marino}, {S{\'a}nchez-Menguiano},
  {Garc{\'{\i}}a-Benito}, {Mast}, {Mendoza}, {Papaderos}, {Ellis}, {Galbany},
  {Kehrig}, {Monreal-Ibero}, {Gonz{\'a}lez Delgado}, {Moll{\'a}}, {Ziegler},
  {de Lorenzo-C{\'a}ceres}, {Mendez-Abreu}, {Bland-Hawthorn}, {Bekerait{\.e}},
  {Roth}, {Pasquali}, {D{\'{\i}}az}, {Bomans}, {van de Ven}, \&
  {Wisotzki}}]{sanchez15}
{S{\'a}nchez}, S.~F., {P{\'e}rez}, E., {Rosales-Ortega}, F.~F., {et~al.} 2015,
  \aap, 574, A47

\bibitem[{{Sanders} {et~al.}(2012){Sanders}, {Soderberg}, {Levesque}, {Foley},
  {Chornock}, {Milisavljevic}, {Margutti}, {Berger}, {Drout}, {Czekala}, \&
  {Dittmann}}]{sanders12}
{Sanders}, N.~E., {Soderberg}, A.~M., {Levesque}, E.~M., {et~al.} 2012, \apj,
  758, 132

\bibitem[{{Scargle} {et~al.}(2013){Scargle}, {Norris}, {Jackson}, \&
  {Chiang}}]{scargle13}
{Scargle}, J.~D., {Norris}, J.~P., {Jackson}, B., \& {Chiang}, J. 2013, \apj,
  764, 167

\bibitem[{Silverman(1986)}]{silverman86}
Silverman, B.~W. 1986, Density Estimation for Statistics and Data Analysis
  (Chapman and Hall)

\bibitem[{{Sim{\'o}n-D{\'{\i}}az} \& {Stasi{\'n}ska}(2011)}]{simondiaz11-orion}
{Sim{\'o}n-D{\'{\i}}az}, S., \& {Stasi{\'n}ska}, G. 2011, \aap, 526, A48+

\bibitem[{{Stasi{\'n}ska}(2002)}]{stasinska02}
{Stasi{\'n}ska}, G. 2002, ArXiv Astrophysics e-prints, astro-ph/0207500

\bibitem[{{Stasi{\'n}ska}(2010)}]{stasinska10}
{Stasi{\'n}ska}, G. 2010, in IAU Symposium, Vol. 262, IAU Symposium, ed. G.~R.
  {Bruzual} \& S.~{Charlot}, 93--96

\bibitem[{{Tremonti} {et~al.}(2004){Tremonti}, {Heckman}, {Kauffmann},
  {Brinchmann}, {Charlot}, {White}, {Seibert}, {Peng}, {Schlegel}, {Uomoto},
  {Fukugita}, \& {Brinkmann}}]{tremonti04}
{Tremonti}, C.~A., {Heckman}, T.~M., {Kauffmann}, G., {et~al.} 2004, \apj, 613,
  898

\bibitem[{{Vanderplas} {et~al.}(2012){Vanderplas}, {Connolly}, {Ivezi{\'c}}, \&
  {Gray}}]{astroml}
{Vanderplas}, J., {Connolly}, A., {Ivezi{\'c}}, {\v Z}., \& {Gray}, A. 2012, in
  Conference on Intelligent Data Understanding (CIDU), 47 --54

\bibitem[{{Vasyliunas}(1968)}]{vasyliunas68}
{Vasyliunas}, V.~M. 1968, \jgr, 73, 5810

\bibitem[{{Zaritsky} {et~al.}(1994){Zaritsky}, {Kennicutt}, \&
  {Huchra}}]{zaritsky94}
{Zaritsky}, D., {Kennicutt}, Jr., R.~C., \& {Huchra}, J.~P. 1994, \apj, 420, 87

\end{thebibliography}

\end{document}